\documentclass[]{aastex701}
\usepackage{soul}
\usepackage{color}
\usepackage{subcaption}  
\begin{document}

\title{Pervasive Cavity-Ring Structure for Star Formation in Dwarf Irregular Galaxies}

\author[0000-0002-1723-6330]{Bruce G. Elmegreen}\affiliation{Katonah, NY 10536, USA}
\email{belmegreen@gmail.com}  

\author[0000-0002-3322-9798]{Deidre A. Hunter}\affiliation{Flagstaff, AZ 86001, USA}
\email{deidreahunter@gmail.com}  

\begin{abstract}
Unsharp-mask images of HI emission from 36 dwarf irregular (dIrr) galaxies illustrate star formation in dispersed clouds and on the rims of large cavities. The cavities can extend for a radial scalelength and typically have circular or slightly sheared forms. The average surface density of cloud peaks is $\sim20\;M_\odot$ pc$^{-2}$, and, combined with their average FUV star formation rate, suggests a gas consumption time of $\sim3.2$ Gyr. Vertical hydrostatic equilibrium calculations for 24 of these dIrrs give a typical scale height of $\sim400$ pc, which combines with the gas and star formation surface densities to suggest an efficiency per free fall time of $\sim1$\%. These values are comparable to those in the molecular clouds of spiral galaxies, suggesting the primary difference between clouds is the presence of CO at higher metallicity in the spirals. $U-B$ color images of the dIrrs suggest that cavity ages range between $10^7$ and $10^8$ years, with the longer times explaining the common lack of bright OB associations in their centers and their low expansion speeds. Most are circular because the shear time exceeds 100 Myr, although some of the HI has spiral structure. These observations suggest that star formation in dIrrs proceeds slowly in a sequential fashion in dispersed clouds and on the periphery of giant cavities that move and expand during the $\sim50$ Myr supernova era of the previous generation. In contrast, spiral galaxies have shear times $10\times$ shorter and more important stellar dynamics that compresses the gas into filaments.
\end{abstract}

\section{Introduction}
Dwarf irregular (dIrr) galaxies are the most common type of galaxy in the local universe. They are relatively thick, gas-rich with clumpy star formation, and have smooth underlying stellar disks with exponential radial profiles \citep{herrmann16,henkel22,hunter24}. They resemble clumpy galaxies at high redshift because both have a high ratio of gas velocity dispersion to rotation speed, which makes their disks relatively thick and the self-gravitating clumps relatively large \citep{elmegreen09}. Local dIrrs are only $\sim1$\% of the mass of their high-$z$ analogs, however, and also lower in density, as the average dynamical time in a galaxy on the star-formation main sequence tends to scale with the age of the Universe \citep{dekel13}, so the average  density scales with the Universe density. 

Low densities for dIrrs make it difficult to understand how stars form \citep{bolatto19}. They lie at the low end of the Kennicutt-Schmidt relation \citep{kennicutt12} with a steeper slope than spiral galaxies \citep{bigiel08,leroy08,roy09}, indicating much slower star formation per unit gas. Consumption times for HI can be 100 Gyr or more, whereas in the main parts of spirals it is several Gyr  \citep{bigiel10}. Toomre Q values measuring disk stability are relatively high in dIrrs when disk thickness is considered \citep{elmegreen15,bacchini20,aditya23}, explaining the lack of spirals and the resulting lack of star formation processes connected with spirals, such as large-scale gas compression and gravitational collapse \citep[e.g.,][]{elmegreen19}. The outer parts of dIrrs are even more extreme, with midplane gas densities as low as $0.1$ cm$^{-3}$ and gas surface densities dominating stars by a factor of $\sim10$ \citep{elmegreen15}. In these outer parts, star formation rates can be traced down to $\sim10^{-5}\;M_\odot$ pc$^{-2}$ Myr$^{-1}$ at average gas surface densities of $\sim1\;M_\odot$ pc$^{-2}$ \citep{elmegreen17}, which is $\sim10\times$ lower than the conventional gas surface density where molecules become prominent in spiral galaxies \citep{bolatto11,park23,schneider25}.  Low metallicities in dIrrs \citep{annibali22} are also a problem for understanding star formation because they make the cooling time long and dust extinction low \citep{krumholz12}, and they make molecular clouds difficult to observe \citep{bolatto13}.

A clue to star formation in dIrrs is suggested by their morphologies. Many have large holes or cavities in their HI emission \citep[for an extreme example, see][]{simpson05}, surrounded by rings of denser HI where star formation takes place.  One of the early examples of this was in the LMC, where \cite{shapley63}, \cite{westerlund66} and \cite{meaburn80} found several large HI cavities and star-formation rings, the largest being Constellation III, or LMC4, with a diameter of $\sim1$ kpc. An examination of the possible energy sources for this cavity \citep{efremov98} suggested that a faint grouping of A-, B- and M-type supergiant stars in the center, along with other stars that formed with them but have since exploded, pressurized the gas for $\sim30$ Myr, with star formation beginning along the periphery after 14 Myr. This central association of evolved stars in LMC4 is not obvious, and would be difficult to see at a larger distance. Indeed, many of the largest HI cavities in dIrrs have no prominent stellar energy sources inside of them \citep{stewart00,silich06,warren11}.  

Among dIrrs, Holmberg II (called DDO 50 in what follows) at 3.4 Mpc distance was the first outside the Local Group to be seen with a large-scale cavity/rim structure in the HI  \citep{puche92}. Most of its star formation is in dense gas at the edges of the cavities, suggesting a time sequence of triggering \citep{egorov17}. Other examples are IC 2574 \citep{walter98,walter99,stewart00,egorov14}, IC10 \citep{wilcots98}, IC 1613 \citep{silich06}, DDO47 \citep{walter01}, Haro 14 \citep{cairos17a}, Tololo 1937-423 \citep{cairos17b}, NGC 6822 \citep{deblok00}, and Holmberg I \citep[called DDO 63 in what follows,][]{ott01,egorov18}.  A systematic investigation of HI cavities and triggered star formation in the Local Irregulars That Trace Luminosity Extremes The HI Nearby Galaxy Survey \citep[``LITTLE THINGS;''][]{hunter12} used HI expansion velocities and star formation rates to suggest that star formation could power the cavities \citep{pokhrel20}, a conclusion made by earlier studies as well \citep[e.g.,][]{stewart00,vorobyov05,warren11}.  This process does not exclude other possibilities, such as turbulence coupled with thermal and gravitational instabilities \citep{dib05}.  Simulations of dIrrs in, for example, \cite{kawata14} and \cite{lahen19} illustrate well the process of star formation along the rims of giant cavities.  \cite{zhao24} simulate this process in spiral galaxies.

Here we study the juxtaposition of HI and FUV emission, along with $U-B$ colors, in 36 dIrrs from the LITTLE THINGS survey.  We highlight HI features with an unsharp-mask (USM) technique (Sect. \ref{sect:enhance}) and compare the HI and FUV images to see where star formation has occurred relative to the HI structure (Sect. \ref{sect:juxtapose}). We introduce a classification of HI cavity morphology (Sect. \ref{cavitytypes}), and another classification of the morphology of FUV with HI (Sect. \ref{sect:juxtapose}). The properties of the peak HI emission regions, which often occur along the rims of the cavities, are determined and discussed in the context of star formation rates (Sect. \ref{sect:SFR}) and thresholds (Sect. \ref{sect:peaks}), and in terms of the ages and colors of regions inside the cavities (Sect. \ref{sect:UB}). The properties of the cavities themselves are also discussed in the context of their ages and relative shear times (Sect. \ref{sect:shear}). A conclusion in Section \ref{sect:conclude} outlines the proposed process of star formation in most of these dIrrs and compares it with star formation in spiral galaxies. 

\begin{center}
\begin{deluxetable}{lccccccc}
\tablenum{1} \tablecolumns{8} \tablewidth{315pt} \tablecaption{The Galaxy Sample \label{properties} } 
\tablehead{
\colhead{} & 
\colhead{D} & 
\colhead{$M_{\rm V}$} & 
\colhead{B-V} &
\colhead{E(B-V)\tablenotemark{a}} &
\colhead{$M_{\rm stars}$} &
\colhead{HI Cavity} &
\colhead{FUV-HI}\\
\colhead{} & 
\colhead{Mpc} & 
\colhead{mag.} & 
\colhead{mag.} &
\colhead{mag.} &
\colhead{$\times10^7\;M_\odot$} &
\colhead{Morph.} &
\colhead{Morph.} 
}
\startdata
CVnIdwA   & $  3.6\pm 0.08$ & $ -12.37\pm 0.08$ & $   0.21\pm 0.11$ & 0.055 & $  0.28\pm  0.14$ & CC & FC, FB        \\
DDO43     & $  7.8\pm 0.80$ & $ -15.07\pm 0.02$ & $   0.31\pm 0.03$ & 0.105 & $  4.35\pm  2.01$ & DC, OC & FC, FB    \\
DDO46     & $  6.1\pm 0.40$ & $ -14.67\pm 0.01$ & $   0.31\pm 0.02$ & 0.103 & $  3.01\pm  1.38$ & DC, OC & FB        \\
DDO47     & $  5.2\pm 0.60$ & $ -15.46\pm 0.01$ & $   0.29\pm 0.02$ & 0.073 & $  5.98\pm  2.69$ & DC, S & FG, FC     \\
DDO50     & $  3.4\pm 0.05$ & $ -16.61\pm 0.00$ & $   0.22\pm 0.00$ & 0.073 & $ 14.34\pm  5.90$ & DC, S & NFC, FG    \\
DDO52     & $ 10.3\pm 0.80$ & $ -15.44\pm 0.03$ & $   0.39\pm 0.04$ & 0.080 & $  7.51\pm  3.90$ & DC & FG, FB        \\
DDO53     & $  3.6\pm 0.05$ & $ -13.84\pm 0.01$ & $   0.41\pm 0.01$ & 0.075 & $  1.78\pm  0.92$ & DC, OC, CC & FG, FC\\
DDO63     & $  3.9\pm 0.05$ & $ -14.78\pm 0.02$ & $   0.20\pm 0.02$ & 0.062 & $  2.54\pm  1.03$ & CC & NFC, FG       \\
DDO69     & $  0.8\pm 0.04$ & $ -11.67\pm 0.01$ & $   0.29\pm 0.01$ & 0.050 & $  0.18\pm  0.08$ & DC, CC & NFC, FG   \\
DDO70     & $  1.3\pm 0.07$ & $ -14.10\pm 0.00$ & $   0.36\pm 0.00$ & 0.063 & $  2.01\pm  0.98$ & DC, OC & FG, FC    \\
DDO75     & $  1.3\pm 0.05$ & $ -13.91\pm 0.01$ & $   0.19\pm 0.01$ & 0.068 & $  1.12\pm  0.45$ & CC & FB, FC        \\
DDO87     & $  7.7\pm 0.50$ & $ -14.98\pm 0.02$ & $   0.45\pm 0.03$ & 0.050 & $  5.69\pm  3.15$ & DC, OC & NFC, FB   \\
DDO101    & $  6.4\pm 0.50$ & $ -15.01\pm 0.01$ & $   0.61\pm 0.02$ & 0.058 & $  8.54\pm  5.65$ & CC, OC & FB, FC    \\
DDO126    & $  4.9\pm 0.50$ & $ -14.85\pm 0.01$ & $   0.29\pm 0.02$ & 0.050 & $  3.37\pm  1.51$ & DC & FG, FC, FB    \\
DDO133    & $  3.5\pm 0.20$ & $ -14.75\pm 0.01$ & $   0.39\pm 0.01$ & 0.053 & $  3.90\pm  1.97$ & DC & FG, FB        \\
DDO154    & $  3.7\pm 0.30$ & $ -14.19\pm 0.01$ & $   0.30\pm 0.03$ & 0.058 & $  1.91\pm  0.87$ & DC, S & FG, FC     \\
DDO167    & $  4.2\pm 0.50$ & $ -12.98\pm 0.04$ & $   0.19\pm 0.05$ & 0.050 & $  0.48\pm  0.20$ & DC & FC, FB        \\
DDO168    & $  4.3\pm 0.50$ & $ -15.72\pm 0.00$ & $   0.37\pm 0.01$ & 0.050 & $  9.20\pm  4.54$ & OC & FG, FC        \\
DDO187    & $  2.2\pm 0.07$ & $ -12.68\pm 0.01$ & $   0.30\pm 0.02$ & 0.050 & $  0.47\pm  0.22$ & OC & FG, FB        \\
DDO210    & $  0.9\pm 0.04$ & $ -10.88\pm 0.01$ & $   0.50\pm 0.01$ & 0.085 & $  0.15\pm  0.09$ & CC & FC, FB        \\
DDO216    & $  1.1\pm 0.05$ & $ -13.72\pm 0.00$ & $   0.66\pm 0.00$ & 0.073 & $  2.97\pm  2.08$ & OC & NFC, FG, FB   \\
F564-V3   & $  8.7\pm 0.70$ & $ -13.97\pm 0.03$ & $   0.37\pm 0.04$ & 0.068 & $  1.84\pm  0.93$ & DC, OC & FG, FB    \\
IC1613    & $  0.7\pm 0.05$ & $ -14.60\pm 0.00$ & $   0.43\pm 0.01$ & 0.055 & $  3.76\pm  1.99$ & DC, CC, OC & NFC, FG\\
LGS3      & $  0.7\pm 0.08$ & $  -9.74\pm 0.03$ & $   0.63\pm 0.05$ & 0.085 & $  0.07\pm  0.05$ & CC & FB            \\
M81dwA    & $  3.5\pm 0.20$ & $ -11.67\pm 0.06$ & $   0.27\pm 0.08$ & 0.072 & $  0.17\pm  0.08$ & CC & FC, FB        \\
NGC1569   & $  3.4\pm 0.20$ & $ -18.24\pm 0.00$ & $   0.25\pm 0.00$ & 0.558 & $ 69.02\pm 29.29$ & DC & FC, FB        \\
NGC2366   & $  3.4\pm 0.30$ & $ -16.79\pm 0.00$ & $   0.30\pm 0.00$ & 0.093 & $ 20.59\pm  9.28$ & DC & NFC, FG       \\
NGC3738   & $  4.9\pm 0.50$ & $ -17.12\pm 0.00$ & $   0.42\pm 0.00$ & 0.050 & $ 37.52\pm 19.67$ & OC & FC, FB        \\
NGC4163   & $  2.9\pm 0.04$ & $ -14.44\pm 0.01$ & $   0.49\pm 0.01$ & 0.050 & $  3.75\pm  2.14$ & OC & FC, FB        \\
NGC4214   & $  3.0\pm 0.05$ & $ -17.63\pm 0.00$ & $   0.37\pm 0.00$ & 0.050 & $ 52.77\pm 25.89$ & DC, OC, S& NFC, FG \\
SagDIG    & $  1.1\pm 0.07$ & $ -12.46\pm 0.00$ & $   0.39\pm 0.01$ & 0.188 & $  0.48\pm  0.24$ & CC & FC, FB        \\
WLM       & $  1.0\pm 0.07$ & $ -14.39\pm 0.00$ & $   0.41\pm 0.00$ & 0.068 & $  2.95\pm  1.53$ & OC & FG            \\
Haro29    & $  5.9\pm 0.30$ & $ -14.66\pm 0.00$ & $   0.26\pm 0.01$ & 0.050 & $  2.63\pm  1.13$ & BCD & FB, FC       \\
Haro36    & $  9.3\pm 0.60$ & $ -15.91\pm 0.00$ & $   0.39\pm 0.01$ & 0.050 & $ 11.42\pm  5.76$ & BCD & FG           \\
Mrk178    & $  3.9\pm 0.50$ & $ -14.11\pm 0.01$ & $   0.36\pm 0.01$ & 0.050 & $  2.04\pm  1.00$ & CC & FC, FB        \\
VIIZw403  & $  4.4\pm 0.07$ & $ -14.27\pm 0.01$ & $   0.28\pm 0.01$ & 0.073 & $  1.93\pm  0.85$ & BCD & FC, FB       \\
\enddata
\tablenotetext{a}{Foreground plus internal color excess.}
\end{deluxetable}
\end{center}

\begin{center}
\begin{deluxetable}{lcl}
\tablenum{2} \tablecolumns{3} \tablewidth{280pt} \tablecaption{HI and FUV cavity Categories \label{types} } 
\tablehead{Galaxy Cavity-Rim Type &\colhead{Example}&
\colhead{Description}
}
\startdata
DC & DDO 133 & ``Distributed Cavities;'' HI cavities cover the\\
&& star-forming disk\\
OC & DDO 70 & ``Outer Cavities;'' large cavities in the outer disk\\
CC & CVnIdwA & ``Central Cavity;'' large cavity in the center\\
BCD & VIIZw403 & ``Blue Compact Dwarf;'' standard definition of\\
&& BCD, having no large HI cavities\\
\hline\hline
\vspace{0.05in}
Galaxy FUV-Cavity Type&Example&Description\\
\hline
FG & DDO 133 & ``FUV-Gas;'' FUV at cloud peak\\
FB & DDO 46 & ``FUV-Blister;'' FUV at cloud edge \\
FC & CVnIdwA & ``FUV-Cavity;'' FUV fills an HI cavity\\ 
NFC & DDO50 & ``No-FUV-Cavity;'' A cavity with little FUV\\
S & DDO 47 & ``Spiral;'' Spiral structure from elongated cavities\\
\enddata
\end{deluxetable}
\end{center}

\section{Image Enhancement for Dwarf Irregular Galaxies}
\label{sect:enhance}
\subsection{Data}
\label{subsectiondata}

The galaxies studied here are listed in Table \ref{properties}. The distances are from \cite{hunter12}. The stellar masses are based on photometry in \cite{hunter06}, using $M_{\rm B}$, $M_{\rm V}$ and mass-to-light ratios from $B-V$ as in \cite{herrmann16}. 

The internal extinctions assume $E(B-V)=0.05$ mag., which was added to the foreground value that is usually a few hundredths, except for NGC 1569 \citep{schlafly11}.  This internal extinction comes from \cite{hunter99}, who used the H$\alpha$/H$\beta$ ratio to determine E(B-V) in 189 HII regions of 65 Im dwarf galaxies. From the galaxy averages of measured E(B-V), they subtracted the foreground reddening obtained from \cite{burstein84} to yield the internal reddening in the HII regions. The average internal reddening in the HII regions was E(B-V)$=0.12$ mag.. Here we adopted roughly half of this value to represent internal reddening not directly associated with star-forming regions. The extinction is so low in dIrrs that slight uncertainties will not affect the results of this paper.

The HI maps used here are a homogeneous sample from our VLA survey\footnote{The VLA is a facility of the National Radio Astronomy Observatory (NRAO), which is a facility of the National Science Foundation operated under cooperative agreement by Associated Universities, Inc. These data were taken during the upgrade of the VLA to the Expanded VLA, now JVLA.} and have NA-weighted beam sizes ranging from $39.8^{\prime\prime}$ for SagDIG to $8.3^{\prime\prime}$ for DDO 69. The SagDIG beam is unusually large; the next largest beam size is $21.3^{\prime\prime}$ for NGC 4163. Pixel sizes are $1.5^{\prime\prime}$ except for DDO 216 and SagDIG, which have $3.5^{\prime\prime}$ pixels. The HI images are from the combination of B, C, and D arrays, with integration times generally about 12, 6, and 2 hours, respectively \citep{hunter12}. 

FUV images are from GALEX \citep{morrissey07}. These images have resolutions of $4^{\prime\prime}$ and pixel sizes of $1.5^{\prime\prime}$. Broadband optical images are used to determine color ages, particularly from $U-B$ color, which is most sensitive to young ages. Most of them were obtained by D.A.H. from 1997 to 2002 with the Lowell Observatory 1.1-m Hall Telescope or the Lowell Observatory 1.8 m Perkins Telescope. They were described in \cite{hunter06} and are available at the National Radio Astronomical Observatory LITTLE THINGS web site\footnote{https://science.nrao.edu/science/surveys/littlethings}, along with the HI data. Ten of the galaxies considered here use ultra-deep U- and B-band images from a recent survey \citep{hunter25}.  They were obtained by D.A.H. from 2014 to 2022 with Lowell Observatory’s 4.3-m Lowell Discovery Telescope and Large Monolithic Imager (LMI).  Sky subtraction, foreground extinction corrections, and other conventional processing are discussed in the survey papers.

\begin{figure*}
\begin{center}
\includegraphics[width=18cm]{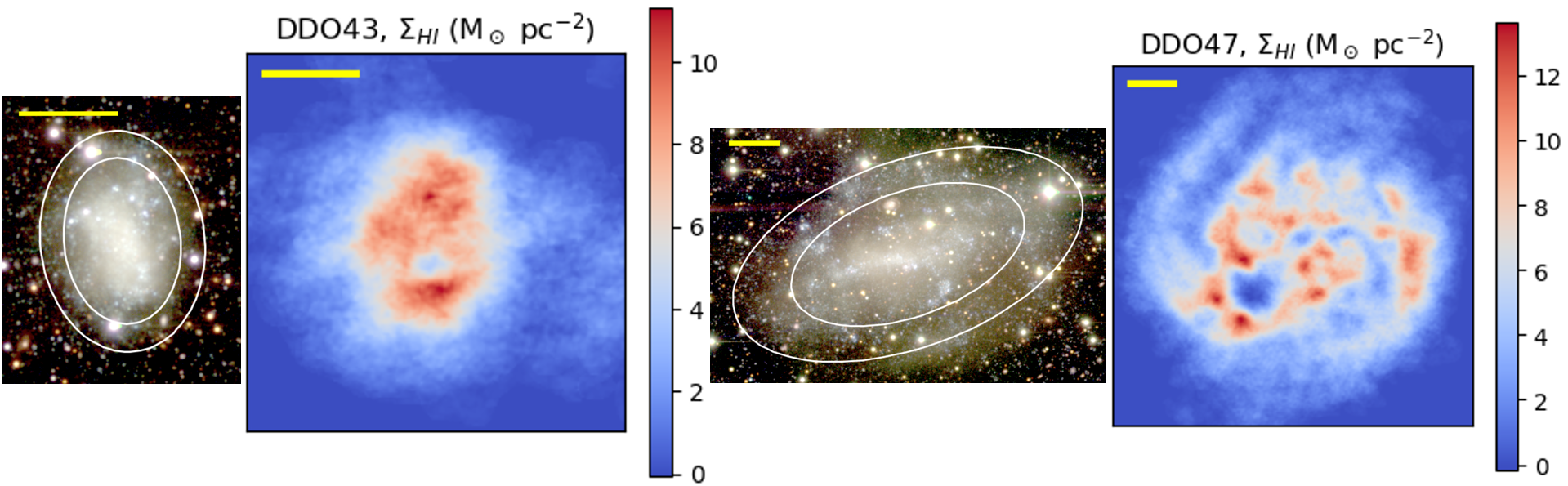}  
\caption{Two examples of dIrr galaxies with ultra-deep images made from U, B, and V bands on the left and HI MOM0 maps on the right. The ellipses on the optical images are at 26 mag arcsec$^{-2}$ and 29 mag arcsec$^{-2}$ as measured in V-band. The color scale on the HI images is in units of $M_\odot$ pc$^{-2}$. There are features in the HI image that are not apparent in the optical images, such as the HI cavities and the gaseous spiral structure in the outer part of DDO 47. The yellow scale bars represent 1 arcmin. 
}
\label{twoexamples}
\end{center}
\end{figure*}

Figure \ref{twoexamples} shows two examples with ultra-deep color-composites made from U, B, and V band images on the left and the total HI (MOM0) map on the right.  North is up in all of the images throughout this paper. The scale bar in yellow represents 1 arcmin and the units of the color scale for HI is $M_\odot$ pc$^{-2}$, where the multiplicative factor of 1.36 has been used to convert the HI column density into a total gas column density, including He and heavy elements. Note that these are projected surface densities, not corrected for galaxy inclination. The ellipses on the optical images are at V-band surface brightnesses of 26 and 29 mag arcsec$^{-2}$. The detected gas extends further out than the largest ellipse in each case and is typically visible in the figures down to $\Sigma_{\rm gas}\sim1\;M_\odot$ pc$^{-2}$. There is clearly structure in the HI that is not present in the optical images, including the big HI cavity in DDO 43, which has no obvious optical counterpart, and the even deeper HI cavity in DDO 47 along with its gaseous spiral arms. \\

\subsection{Unsharp Masks Reveal Four HI Cavity Types}
\label{cavitytypes}

A few of the dIrr in our survey have obvious cavity/rim structures in HI and star formation, as cited in the Introduction. Others do not show this structure in total emission HI maps because of a more uniform component of additional emission. To remove the uniform component, we performed an unsharp-mask (USM) operation on the moment-zero (MOM0, total emission) HI map. The USM subtracts a blurred version of the image from the original image, leaving only the local structure without the smooth component. The blurred image was made by convolving the original image with a Gaussian that has a dispersion equal to half the radial scale length measured in V-band \citep[from][]{hunter21}. This scaling with galaxy size emphasizes a certain common relative scale, as the USM highlights all features equal to or smaller than the blurring scale. With half the disk scalelength for the blurring scale, the main effect of this operation is to remove the exponential disk and other galaxy-wide emission.

\begin{figure*}
\begin{center}
\includegraphics[width=18cm]{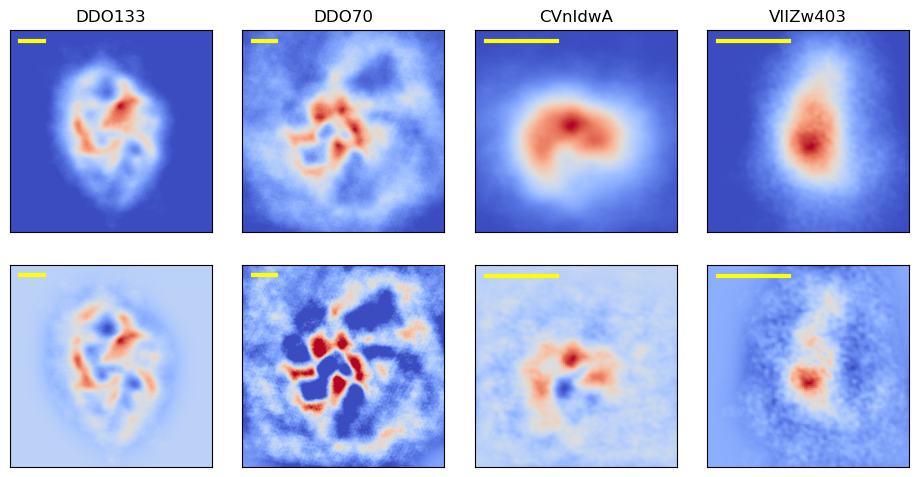}  
\caption{(top) HI emission from galaxies showing four different HI morphologies from left to right: large dispersed cavities (DC) throughout the main disk, large outer cavities (OC) beyond the detected FUV emission, possibly in addition to DC; a large central cavity (CC), and BCDs galaxies, which have centrally concentrated HI and no obvious cavities. 
The color scale ranges from blue at low emission to red at high emission. The limiting projected surface densities for HI in the top panels are, from left to right: 0 to 15.1, 0 to 17.6, 0 to 29.4, and 0 to 24.5, in units of $M_\odot$ pc$^{-2}$. (bottom) Unsharp-mask images of the same four galaxies, made by subtracting a Gaussian blur from the original HI image using a Gaussian dispersion equal to one-half of the V-band radial scale length of the galaxy. The yellow scale bars represent 1 arcmin.
}
\label{fourtypes}
\end{center}
\end{figure*}

After the USM, almost all of the galaxies are seen to contain large HI cavities with smaller HI clouds and filaments around the rims. We identified four main configurations for this structure, as shown in Figure \ref{fourtypes}, where total HI emission is in the top panel and the USM image is in the bottom panel. On the left is DDO 133, which has large HI cavities {\it distributed} all over its star-forming disk;  we refer to this type as DC, for ``distributed cavities''. Next is DDO 70, which has cavities in the {\it outer} disk, beyond the FUV emission, designated as type ``OC''. The third type is shown by CVnIdwA in the third panel, which has a large cavity in the {\it center} (type ``CC''). The fourth has no large HI cavities but a central HI concentration instead; these are Blue Compact Dwarfs (BCDs) in our survey, as exemplified by VIIZw403 in the figure. All of the images are plotted on a linear intensity scale with color from blue to red spanning the total range, as given in the caption. For the USM images, the fainter parts have negative values.   These HI cavity types with brief descriptions are summarized in Table \ref{types}.  Representative and dominant types for each galaxy are in Table \ref{properties}.

Other figures in this paper show the rest of our sample and indicate the HI peaks and cavities with cyan and yellow contour lines. A complete set of images is in the Appendix. For some galaxies, two cavity types are present, such as DC and OC, or CC and OC, which designate cases, respectively, where there are cavities in both the main optical disk and the far-outer HI disk (e.g., DDO 43 and DDO 46 in Fig. \ref{appendix1}, DDO 70 in Fig. \ref{fourtypes}) and cavities just in the center and far-outer parts (e.g., DDO 101 in Fig. \ref{appendix3}).  DDO 53 (Fig. \ref{appendix2}) has all cavity types, DC, OC, and CC because there is a deep cavity in the center, shallower cavities throughout the main disk and further shallow cavities in the far-outer disk.

Some dIrrs also have clear spiral structure in the USM HI emission (e.g., DDO 47 in Fig. \ref{twoexamples}, DDO 50 in Fig. \ref{appendix1}, NGC 4214 in Fig. \ref{appendix5}), which is not obvious in the total HI or optical emission. We designate these as type ``S'' and identify specific cases in Tables \ref{properties} and \ref{types}. 

\begin{figure*}
\begin{center}
\includegraphics[width=18cm]{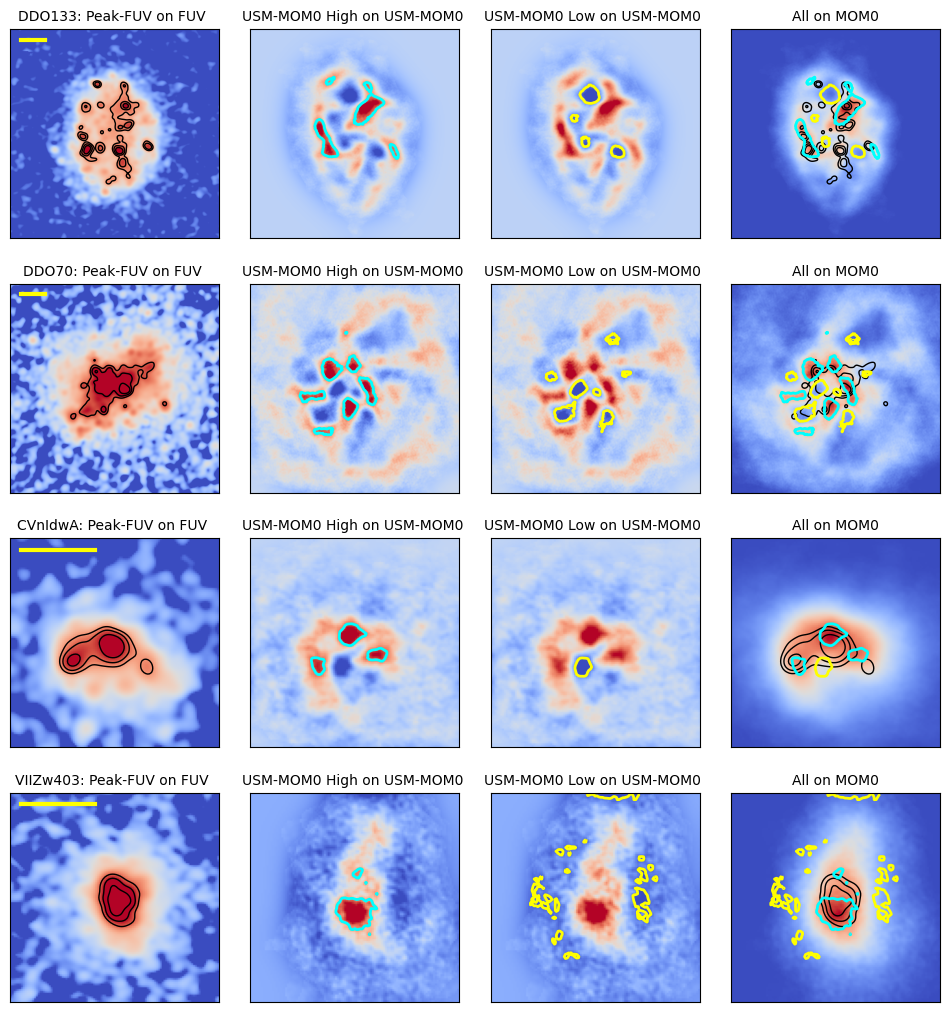}  
\caption{(Rows) Galaxies from Figure \ref{fourtypes} with image processing to reveal morphologies of FUV emission relative to HI structure. (Columns from left to right: the FUV image with contours of FUV emission superposed; the USM image of HI emission (from Fig. \ref{fourtypes}, bottom) with cyan contours for the peak regions; the USM image of HI with yellow contours for the cavities, and the full HI image with all the other contours superposed. DDO 133 shows FUV emission mostly at the positions of the HI peaks with a few regions where FUV extends into the HI cavities; DDO 70 has FUV emission in large regions free of HI; CVnIdwA has FUV and HI mostly coincident as in DDO 133 and the large central HI cavity has relatively little FUV, and VIIZw403 has FUV mostly on the large central cloud with some extending north to several small clouds.
}
\label{FUV-USM}
\end{center}
\end{figure*}

\subsection{FUV - Cavity Juxtapositions and FUV Morphological Types}
\label{sect:juxtapose}

Figure \ref{FUV-USM} compares the USM images of HI with the FUV images for these same four galaxies. All are presented on a linear scale of intensity varying from blue to red.  On the left are FUV images with contours at three intensity values equally spaced in the log of intensity as 
\begin{equation}
\log(I_{\rm FUV,contour})=\log(I_{\rm FUV,min})+A[\log(I_{\rm FUV,max})-\log(I_{\rm FUV,min})]
\label{FUVcontour}
\end{equation}
for $A=0.7$, 0.8 and 0.9 and minimum and maximum intensity values $I_{\rm FUV,min}$ and $I_{\rm FUV,max}$.  The second panel has the USM version of the total hydrogen emission with one cyan contour for the peaks given by
\begin{equation}
I_{\rm USM-HI,peak}=10^{0.85\log(I_{\rm USM-HI,max})}.
\label{hicontour}
\end{equation}
The third from the left has the same USM image with one yellow contour around the cavities given by the negative value,
\begin{equation}
I_{\rm USM-HI,cavity}=-10^{0.85\log(-I_{\rm USM-HI,min})}
\label{hicavity}
\end{equation}
The fourth panel combines these three images by plotting $I_{\rm FUV,contour}$ on top of the total HI image, with USM HI peaks and cavities enclosed by cyan and yellow contours. This fourth panel shows how the FUV corresponds to the USM HI features.  The constants in these equations were chosen to highlight the main features on the FUV and USM HI images after examining a wide range of values. The qualitative results of this paper do not change if these constants vary by$\sim10$\%.  The cavity types in Table \ref{properties} are not always present in the distribution of yellow contours, which outline primarily the deepest cavities.

The star formation rates corresponding to the three FUV contours are given in Appendix Table \ref{thresholds}. The cavity and peak contours in the USM HI images do not have particular values in $\Sigma_{\rm HI}$ because the USM image is made by subtracting a different amount from each HI-MOM0 pixel. Thus, there is no one-to-one correspondence between a yellow contour value and $\Sigma_{\rm HI}$. To estimate characteristic values at the cavity contours, we determined the averages and standard deviations of the projected $\Sigma_{\rm HI}$ (corrected for He and heavy elements) for pixels with USM values within 1\% of the USM thresholds given by equations (\ref{hicontour}) and (\ref{hicavity}). These are also in Table \ref{thresholds}. The last entry is the average value of projected $\Sigma_{\rm HI}$ inside the peak contour given by equation (\ref{hicontour}), again corrected for He and heavy elements. 

Figure \ref{FUV-USM} illustrates several points about the juxtaposition of FUV and HI that appear in the other galaxies too (see the Appendix). DDO 133 has most of the FUV where the HI peaks and very little in the cavities, although there are slight excursions of FUV into low-HI regions, such as in the northwest and southwest. These slight excursions sit on cavity edges and suggest that sometimes the pressure to make or re-clear a cavity comes from the side with a blister-like expulsion of gas away from a cloud \citep{israel78,tenoriotagle79}. 

DDO 70 in Figure \ref{FUV-USM} is different. It has a large centralized FUV region inside an HI cavity with a ring of six HI peaks around it and an extension of the FUV through the two peaks in the northwest.  It also has large outer-disk cavities without significant FUV emission in them. The similarity of these outer HI cavities to those in the main disk of DDO 70 and in other galaxies studied here suggests they were made by the same processes. These processes are most likely associated with star formation even though the FUV is not visible now. 

Also in Figure \ref{FUV-USM}, CVnIdwA has a large region of FUV emission partially surrounding a deep HI cavity. There is also significant FUV between the HI peaks in the northeast. This may be another example of cavity formation from star formation pressures on the side with slightly older FUV stars between the main peaks representing a previous generation. We will see in Section \ref{sect:UBmaps} when we discuss $U-B$ maps that the cavity in CVnIdwA contains mostly old stars and is surrounded by young stars in the north where the FUV appears, but it also has somewhat young stars encircling it in the south. 

The last example is VIIZw403, which is a BCD galaxy with one prominent, centralized HI cloud offset to the south of the FUV, which is more centralized. The northern half of the galaxy has FUV without much HI, although there is no obvious cavity there in the USM image. There is however a trail of HI emission in the north and low-level HI to the east and west. A small HI peak lies at the northern extent of the bright FUV. 

These images suggest several morphological classes for the juxtaposition of FUV and HI. We will call examples like DDO 133 where significant FUV is at the same position as an HI gas peak type FG, for ``FUV-Gas.''  Good examples of FUV at the edge of a cloud and partially extending into an HI cavity will be called ``FUV-Blister'', or FB. When FUV largely fills an HI cavity, we will use the term FC, for ``FUV-Cavity.'' If a cavity has essentially no FUV, then that is an NFC, or ``No-FUV-Cavity.'' Table \ref{properties} lists the prominent types of FUV-cavity juxtapositions in each galaxy and Table \ref{types} summarizes their descriptions. As for the HI cavity morphological types, there is generally a mixture of FUV-HI types in each galaxy and the types are also a matter of degree. The designations in Table \ref{properties} highlight the clearest examples.  To make these FUV-HI morphological type designations, we use Figures A2-A7 in the Appendix, which show all 36 galaxies  with FUV and USM HI contours superimposed in various way, plus a map of $U-B$ color, which will be discussed in Section \ref{sect:UBmaps}. 

Presumably star formation occurs only in the clouds, but FUV tracks both current and past star formation, back for several tens of Myr as a population ages. The FUV-HI types FG and FB are about where the young stars are now relative to the cloud peaks, i.e., centered or mostly to the side, and the FC and NFC types are about  where the young stars were in the recent past, i.e., inside the current cavities or possibly too old to see there. The U-B colors discussed in Section \ref{sect:UB} are a better measure of region age than the presence or lack of FUV.  These colors will show that stars in the cavities tend to be old and stars near the HI peaks tend to be young. We will see that the cavity age is important because of its connection to the shear time (Sect. \ref{sect:shear}), and the HI peak age is important because of the connection between the star formation rate and HI mass, which gives the star formation efficiency and gas consumption time in the clouds (Sect. \ref{sect:peaks}).

\section{Relative sizes of Clouds, Cavities and Star-Forming Regions}

\begin{figure*}
\begin{center}
\includegraphics[width=16cm]{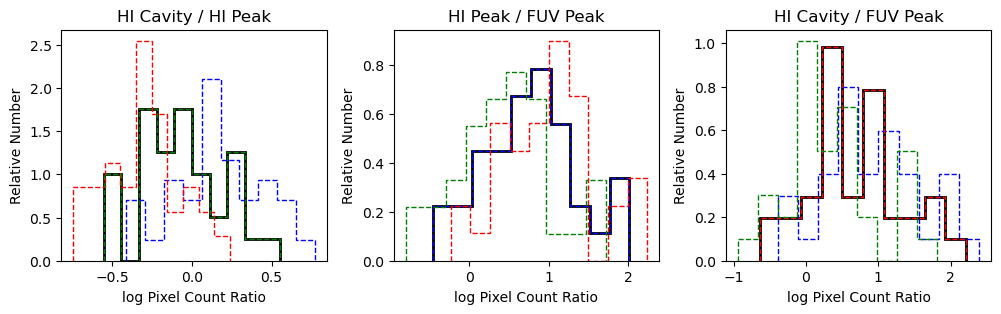} 
\caption{Histograms of the ratios of the numbers of pixels in various regions of the 36 galaxies. These ratios are from the total pixel counts in each galaxy above or below three thresholds.  On the left is the distribution of the ratio of the total HI cavity area to the total HI peak area for all of the galaxies, as defined in the USM images. In the middle is the distribution of the ratio of the USM HI and FUV peak areas. On the right is the distribution of the ratio of the USM HI cavity area to the FUV peak area. The area of brightest star formation is $\sim21$\% that of the cavities defined on the USM images. Dashed lines show variations in these ratios for slightly different definitions of the thresholds (see text).   
}
\label{histratio}
\end{center}
\end{figure*}

The USM HI images reveal cavity/rim structures that are often not recognized in the total HI (MOM0) maps. This structure can dominate the overall appearance of the galaxy in the USM HI images, even when there is no evidence for it in optical or FUV images. To quantify the relative importance of the cavities compared to the HI and FUV peaks that tend to outline the cavity rims, we summed the number of pixels inside all the HI cavities and peaks defined by equations (\ref{hicavity}) and (\ref{hicontour}), and inside all the FUV peaks defined by equation (\ref{FUVcontour}) for each galaxy, and we took the ratios of these sums. These ratios do not depend on inclination and are made on the projected images. Figure \ref{histratio} shows histograms of the log of these ratios. The mean and standard deviation of the log of the ratio of the total HI cavity area to the total HI peak area is $-0.077\pm0.265$, so this ratio is $\sim0.84$ and the cavities occupy about the same area as the HI peaks. This is sensible given the same constants in equations (\ref{hicontour}) and (\ref{hicavity}).  For the ratio of cavity to FUV peak areas, the mean of the log is $0.678\pm0.645$, which corresponds to an area ratio of $\sim4.8$; i.e., most of the FUV is confined to an area of 21\% of the HI cavities.

Uncertainties in these ratios are shown in Figure \ref{histratio} using histograms with dashed lines. These dashed histograms represent the same definitions of ratios but for contour levels that use constants in equations (\ref{FUVcontour})-(\ref{hicavity}) that are successively lower by 5\%, one at a time.  Because the FUV and HI peaks are relatively small and steep, slight variations in the contour values of these peaks correspond to relatively large variations in the area ratios. A lower FUV threshold (green dashed histogram) increases the number of pixels in the FUV peaks and shifts the 2nd and 3rd histograms to the left. A lower USM HI peak threshold (red dashed histogram) includes more HI peak pixels, and shifts the 1st histogram to the left and the 2nd to the right. A less negative USM HI cavity threshold (blue dashed histogram) increases the number of pixels in the cavities and shifts the 1st and 3rd histograms to the right. The three sets of dashed lines tend to be centered around the main histogram, and each has a displacement from the main one that is about one standard deviation.

\section{Star Formation Rate and Gas Surface Density}
\label{sect:SFR}

The FUV counts in the GALEX images, $I_{\rm FUV}$, were converted into areal star formation rates in $M_\odot$ pc$^{-2}$ Myr$^{-1}$ using the equation in \cite{hunter10},
\begin{equation}
\log(\Sigma_{\rm SFR})=\log(I_{\rm FUV})-0.769,
\end{equation}
where the constant 0.769 is for 1.5" pixels; two galaxies, DDO 216 and SagDIG, have 3.5" pixels and the constant for them is 1.505. The Chabrier IMF was assumed using a mass factor of 0.55 compared to the Salpeter function in \cite{kennicutt98} and \cite{hunter10}. 

\begin{figure*}
\begin{center}
\includegraphics[width=8cm]{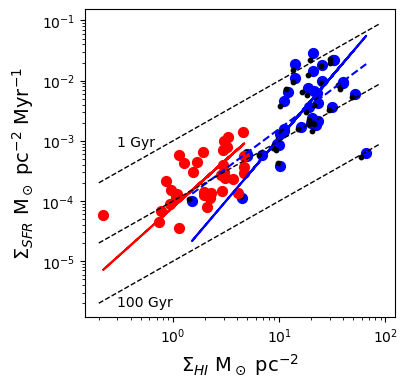} 
\caption{Blue points: the average projected star formation rate per unit area is plotted against the average projected HI surface density (corrected for He and heavy elements) for all regions where the USM HI value exceeds the threshold given by the cyan contours on the USM HI images, from equation (\ref{hicontour}). Red points: Average star formation rate densities versus average HI surface densities for all the regions where FUV is detected. The small black points are like the blue points but with a 5\% lower threshold USM HI intensity. Each galaxy is a distinct point. The blue and red lines are the linear bivariate  (i.e., symmetric) fits to each group of points, while the dashed blue line is a linear fit to only $\Sigma_{\rm SFR}$ versus $\Sigma_{\rm HI}$. The dashed black lines show consumption times of 1 Gyr, 10 Gyr and 100 Gyr, from top to bottom.
}
\label{SFR}
\end{center}
\end{figure*}

Using the FUV images, we calculated the average projected star formation rate per unit area, $<\Sigma_{\rm SFR}>$, in all the regions where the HI intensity exceeds the value chosen to define the USM HI peaks, i.e., where $I_{\rm HI}>I_{\rm USM-HI,peak}$ from equation (\ref{hicontour}). We also determined the average projected surface density of HI there, $<\Sigma_{\rm HI,peaks}>$, multiplied by 1.36 to account for He and heavy elements. The values are plotted as large blue points in Figure \ref{SFR}.  The large red points are average projected star formation rates and HI surface densities for all regions where FUV emission is detected; the values are smaller because most FUV is outside the USM HI peaks. Dashed lines are HI depletion times, $\Sigma_{\rm HI}/\Sigma_{\rm SFR}$, equal to 1 Gyr, 10 Gyr and 100 Gyr. 

A bivariate linear fit correlation for the peaks (large blue points in the figure) is
\begin{equation}
<\Sigma_{\rm SFR}> = (2.08\pm 0.76) <\Sigma_{\rm HI,peaks}> -(5.04\pm 0.92).
\label{linearfit}
\end{equation}
A bivariate fit for all the non-zero FUV regions has a slope of $(1.58\pm0.78)$ and an intercept of $-(4.10\pm 0.23)$, but this is not relevant to star formation because it contains a lot of HI cavity regions.  For 24 LITTLE THINGS galaxies considered recently elsewhere \citep{elmegreen25}, the {\it deprojected} azimuthally averaged relation between $\Sigma_{\rm SFR}$ and $\Sigma_{\rm HI}$ is similar to the {\it projected} relation here for the peaks, with a slope of $(2.18\pm0.55)$ and an intercept of $-(4.86\pm0.26)$. The same bivariate fitting method was used in the two cases.

For these bivariate fits, we first do a least-squares fit of the data to the linear function $y=A_1+B_1x$ for abscissa $x$ and ordinate $y$, and then fit the other direction, $x=A_2+B_2y$. The bivariate fit is $y=A_3+B_3x$, bwhere the value of $A_3$ is taken to be the average of the two intercepts, $A_3=(A_1-A_2/B_2)/2$ and the uncertainty in $A_3$ is taken to be half the full range of the two intercepts, $(A_1+A_2/B_2)/2$. This solution considers that the second fit is the same as $y=x/B_2-A_2/B_2$.  Similarly for the slopes, $B_3$ is the average of the slopes,  $(B_1+1/B_2)/2$ and the uncertainty in $B_3$ is half the range, $(B_1-1/B_2)/2$.  This type of fit gives an equivalent result for either variable as a function of the other, $y(x)$ or $x(y)$, i.e., it assumes no preferred dependent variable and no implied direction of causality. A conventional linear fit to $y(x)$ is shown by the dashed blue line in Figure \ref{SFR}.  It is the same as the direct fit, $y=A_1+B_1x$, which is $<\Sigma_{\rm SFR}> = (1.31\pm0.24)<\Sigma_{\rm HI,peaks}>-(4.12\pm0.16)$.

The points in Figure \ref{SFR} are not a homogeneous sample as the resolution is different for each galaxy. However, we showed in \cite{elmegreen25} for 24 dIrrs that the relationship between HI surface density and star formation surface density (in that case, for deprojected values) was independent of resolution between $\sim 24$ pc and $\sim 424$ pc. The sample of blue points is also inhomogeneous because the values are only for the peak HI regions, and these are defined differently for each galaxy in terms of physical surface density (see Table \ref{thresholds}), although they are defined in the same way in terms of the range of USM HI intensities (equation \ref{hicontour}). 

To determine the sensitivity of the values to the choice of HI peak regions, Figure \ref{SFR} also shows smaller black points that use a 5\% lower threshold HI intensity, i.e., using the constant 0.8075 instead of 0.85 in equation (\ref{hicontour}) to define the peaks inside which both $<\Sigma_{\rm SFR}>$ and $<\Sigma_{\rm HI}>$ are measured.  Each point shifts to slightly lower values in both quantities, but the relationship hardly changes, now becoming $<\Sigma_{\rm SFR}> = (2.05\pm 0.78) <\Sigma_{\rm HI,peaks}> -(4.98\pm 0.92)$ instead of equation (\ref{linearfit}).

Figure \ref{SFR} suggests that although the relationship between star formation rate and gas density is typically steep for dIrr galaxies and other low surface brightness regions \citep[e.g.,][]{roy09}, most of the USM HI peaks in these dIrrs have HI consumption times that are normal for molecular gas in spiral galaxies, i.e., between 1 and 10 Gyr.  We discuss this more in Section \ref{sect:peaks}.

\section{$U-B$ images and region ages}
\label{sect:UB}

$U-B$ color is a more sensitive indicator of age than the presence or lack of FUV emission. Here we discuss how color changes with age for a star-forming region superposed on an old background of stars, and we use this color function to convert a $U-B$ map of two galaxies into age maps. The Appendix contains $U-B$ color maps of all 36 galaxies, but no age maps because they look qualitatively similar to the color maps. 

\subsection{Color versus Age}
\label{sect:UBversustime}

\begin{figure*}
\begin{center}
\includegraphics[width=8cm]{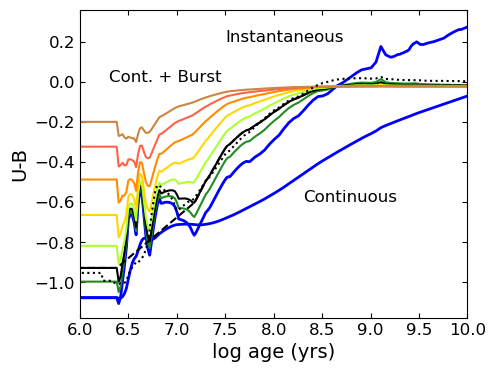} 
\includegraphics[width=7.2cm]{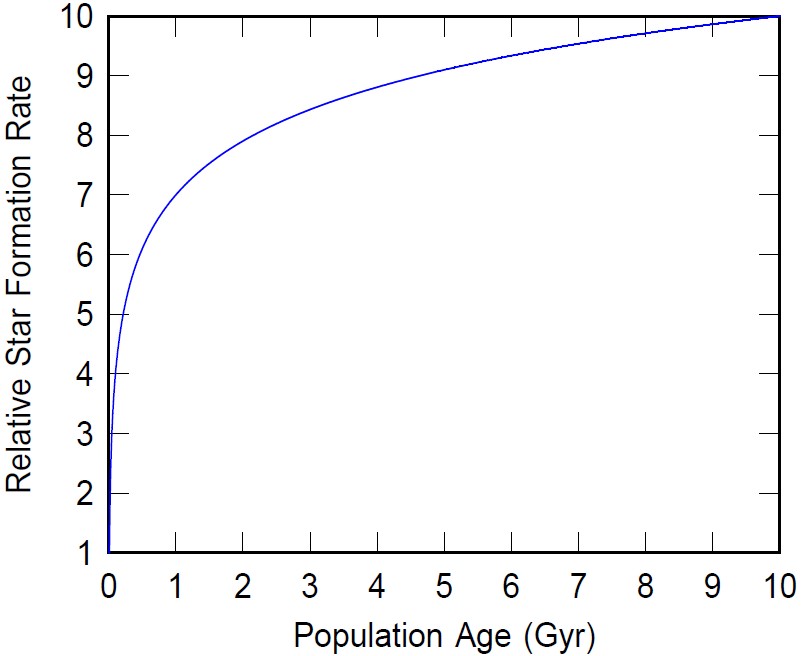} 
\caption{(Left) $U-B$ color versus age for metallicity $Z=0.004$ from \cite{bruzual03}. The lower blue curve is for continuous star formation, the upper blue curve is for an instantaneous burst, and 6 multi-color curves plus a black curve are for bursts with ages on the abscissa superposed on a background of stars that formed by the continuous history of the lower solid blue line. The asymptotic values of the multi-color and black curves equal the value of the continuous curve at large age. The multi-color and black curves differ in the strength of the superposed burst, varying from 2 to 128, top to bottom, in powers of 2, with black being 64. For this black curve, we use a linear interpolation of the $U-B$ color versus age between log-age values of 6.38 and 7.18 (dashed black curve) to smooth over the rapid variations. The dotted black curve is the same as the black solid curve but for metallicity $Z=0.008$. The continuous star formation for the solid blue line and the underlying old population of the superposed bursts has a star formation rate that increases back in time according to equation (\ref{history}) with $F=10$, which is shown on the right.
}
\label{fig:UB}
\end{center}
\end{figure*}

Figure \ref{fig:UB} (left) shows $U-B$ color versus population age for several star formation histories. The blue curve labelled ``continuous'' represents a region with continuous star formation rate following a history where the rate increases logarithmically toward the past as 
\begin{equation}
SFR = F-(F-1)\times{{\log({\rm age,max})-\log({\rm age})}\over{\log({\rm age,max})-\log({\rm age,min})}}.
\label{history}
\end{equation}
The factor $F$ controls how much larger the rate was at the maximum age ``age,max''. The single stellar population models are from \cite{bruzual03} so the minimum and maximum ages are from their tables of U and B magnitudes versus time, namely $\sim10^6$ years to $\sim10^{10}$ years  (the actual range in \cite{bruzual03} is $\log age=5.1$ to 10.3, but the colors do not change much outside the range from 6 to 10).  We found that variations in $F$ do not affect the results significantly because $U-B$ is most sensitive to the recent history. We used $F=10$ here. We also found that the expression for history does not matter much either for the $\sim100$ Myr timescale of interest, but chose an increasing past history anyway because that is realistic.  Figure \ref{fig:UB} (right) shows equation (\ref{history}) with $F=10$. The point of using a star formation history is that the observed young regions today have young stars that are superposed on old stars, so the colors of the composite population depends on the underlying old disk.  For the pixel maps of color in the figures, there is no background disk subtraction as there might be when determining the color of an isolated region, so the pixel values cannot be converted directly to the population color of just the young region. 

The blue curve labelled instantaneous in Figure \ref{fig:UB} (left) is not used but is for reference purposes: it is the color of a single population that ages with time. The multi-color curves and black curve show the variations of $U-B$ with age of single stellar populations that begin at a time of $10^6$ yrs on the abscissa and are superposed on the older population whose U and B magnitudes have the values represented by the large-age point of the continuous population, i.e., after following equation (\ref{history}) for $10^{10}$ years. Note that for large ages of the single population, it becomes so faint that its colors are indistinguishable from the background colors (i.e., the color of the continuous curve at large age). The multi-color and black curves differ by the relative amount of star formation in this recent burst, which varies in powers of 2 as 2, 4, 8, 16 etc. up to 128. This means that the young population which is following the instantaneous color curve by itself has a rate equal to this burst value times the linearly-averaged rate of all the previous star formation.  The weakest burst is the top brown curve and the strongest burst is the bottom green curve. The black curve assumes a burst strength equal to 64 times the average past star formation rate. We consider this to be a good model for our galaxies because star-forming regions are concentrated and have local high star formation rates, although low stellar densities, compared to their blended underlying populations. This choice is not critical, but it does determine the range of colors presented in the $U-B$ maps, i.e., from $U-B=-1$ to $U-B=0$ as discussed above. The portion of the black curve between $\log ({\rm age})=6.38$ and $\log ({\rm age})=7.18$ has been interpolated linearly between the values in \cite{bruzual03} to avoid ambiguous population colors, which are seen as sharp variations in $U-B$ with age in this interval. With the black curve and this dashed interpolation in Figure \ref{fig:UB}, the colors in the $U-B$ maps were converted to approximate ages of the most recent star formation bursts. 

Metallicity affects population color by a relatively small amount compared to age. The multi-color and black curves in Figure \ref{fig:UB} are for metallicities of $Z=0.004$, compared to solar which is $Z=0.02$. The dotted black curve in Figure \ref{fig:UB} uses a metallicity of $Z=0.008$ and a star formation burst factor of 64, i.e., to be compared with the solid black curve. The dotted black curve is slightly lower than the solid black curve at small age and larger at large age, but the difference is not enough to significantly change the inferred age when added to the background population at the same metallicity (as is done in the figure). 

\subsection{Color and Age Maps}
\label{sect:UBmaps}

\begin{figure*}
\begin{center}
\includegraphics[width=14cm]{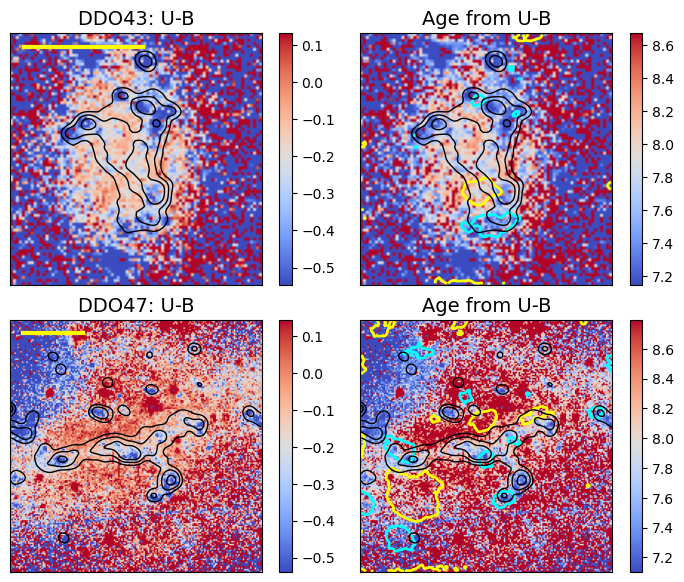} 
\caption{Ultra-deep $U-B$ color map of 2 galaxies are on the left, with color bar giving the values in magnitudes, and age maps based on these $U-B$ colors are on the right, with log age in years indicated on the color bar. Age was determined from color using the black curve in Fig \ref{fig:UB}. The black contours on all four maps outline the peaks of FUV emission, given by equation (\ref{FUVcontour}). The cyan contours on the age maps show the regions of high HI intensity, from equation (\ref{hicontour}), and the yellow contours show the cavities, from Equation (\ref{hicavity}). The star-forming regions indicated by the FUV are clearly bluer in color and younger in age, while the regions between the HI peaks are generally redder and older. Still, there is a lot of intermediate age emission with grayish colors on the age map in regions outside the HI peaks; the ages there are slightly less than $10^8$ years. There are also HI peaks with old ages and no FUV or blue emission. Maps like these can be used to infer the ages of the stellar populations in the cavities where both HI and FUV emissions are weak. 
}
\label{fig:UBimage}
\end{center}
\end{figure*}

Figure \ref{fig:UBimage} shows $U-B$ and age maps of the two galaxies with ultra-deep images in Figure \ref{twoexamples}. The black contours in each image are peak FUV emission, from equation (\ref{FUVcontour}), the cyan contours in the age images are peak USM HI emissions from equation (\ref{hicontour}),  and the yellow contours in the age images define the USM HI cavities, from equation (\ref{hicavity}).  The $U-B$ colors are converted to ages using the black curve in Figure \ref{fig:UB}. The other $U-B$ curves give about the same result but the black curve was chosen because some of the $U-B$ values in other galaxies are close to $-1$, and this implies the most recent bursts have high star formation rates. 

The star-forming regions indicated by FUV contours are also blue in $U-B$ and young in the age maps, as expected. More interesting are the colors and ages in the cavities, i.e., inside the yellow contours. For DDO 43, there are some cavity regions with young to intermediate ages, $\log age \sim7.7$, and other cavities with somewhat old ages, $\log age \sim 8.0$. In DDO 47, most of the cavities are older, $\log age \sim8.5$, with a few small regions having $\log age \sim7.7$. All ages are in years. 

\begin{figure*}
\begin{center}
\includegraphics[width=12cm]{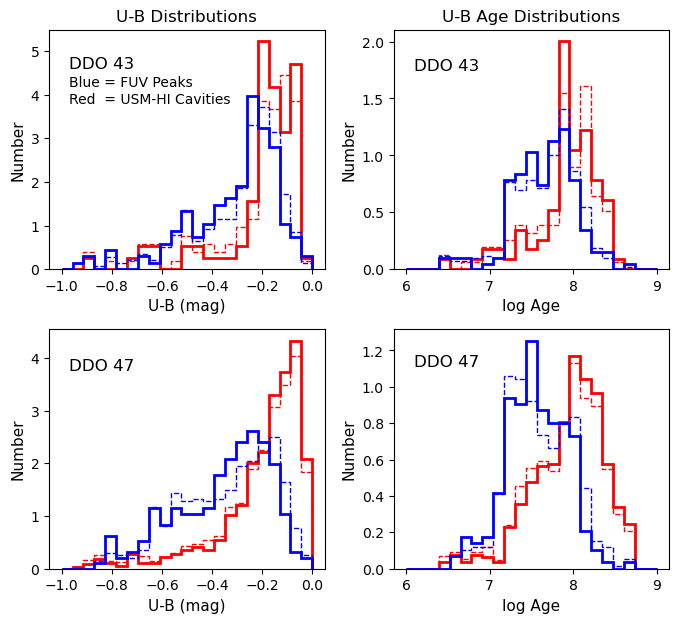} 
\caption{Histograms of $U-B$ color (left) and age (right) for regions (blue lines) where the FUV intensity exceeds the equation (\ref{FUVcontour}) peak value using $A=0.9$ and regions (red lines) where the USM HI intensity is lower than the equation (\ref{hicavity}) cavity value using the constant 0.85. Dotted lines use values of 0.855 and 0.8075 for these thresholds, respectively. The two galaxies are the same as those in Figure \ref{fig:UBimage}.
}
\label{fig:UBimage_hist}
\end{center}
\end{figure*}

The distributions of color and age for these two galaxies are shown in Figure \ref{fig:UBimage_hist} (analogous distributions for all 36 galaxies are shown in Fig. \ref{appendix7}). The solid blue histograms are the distributions of color (left panel) and age (right panel) for regions inside the FUV peaks, defined by $A=0.9$ in equation (\ref{FUVcontour}). The solid red histograms are the distributions of color and age for regions inside the HI cavities, defined by the constant 0.85 in equation (\ref{hicavity}).  The dashed histograms show variations if these intensity thresholds are changed: $A=0.855$ instead of 0.9 for the FUV peaks and 0.8075 instead of 0.85 for the HI cavities. These variations move each distribution toward the distribution with the other color.  A lower threshold for defining the star formation peaks causes slightly older regions to be included, so the color distribution moves toward the red and the age distribution moves toward older. Similarly, a less-deep cavity tends to move the cavity colors bluer and the ages younger. 

The color and age distributions in Figure \ref{fig:UBimage_hist} are about the same for the two galaxies. The log-ages inside the FUV peaks range from $\sim7.1$ to $\sim8$ and the log-ages inside the HI cavities range from $\sim7.8$ to $\sim8.3$. The cavities are clearly older than the FUV peaks, as they should be, with a typical value of $\sim100$ Myr. The FUV peaks are a little older than might be expected for star-forming regions, but this is consistent with Figure \ref{fig:UBimage}, which has some of the highest FUV regions (black contours) extending into the HI cavity regions (yellow contours) where the pinkish color in the age figures corresponds to $\sim100$ Myr ($U-B\sim-0.2$). Star-forming regions that have recently been cleared of gas correspond to the left-hand parts of the red contours, with young ages. We previously noted these cleared regions as horizontal streaks of points in a pixel-by-pixel plot of $\Sigma_{\rm SFR}$ versus $\Sigma_{\rm gas}$ \citep{elmegreen25}.

$U-B$ maps for all 36 galaxies are shown in Figures A2-A7 in the Appendix, with the colorbar for the $U-B$ images in Figure A1. For these $U-B$ figures, most of the galaxies use the broadband images that are on the NRAO LITTLE THINGS web site, which are from \cite{hunter06}, but 10 galaxies use the ultra-deep images from \cite{hunter25}, as indicted by ``Deep'' in the $U-B$ image labels. The image color has the same range for each galaxy, from blue at $U-B=-1.0$ mag to red at $U-B=0$ mag, with a linear progression of $U-B$ between these limits. Thus gray color between blue and red has $U-B=-0.5$ mag. All the contours are superposed on the $U-B$ images:  black for FUV, cyan for USM HI peaks and yellow for USM HI cavities.  The other panels show FUV contours on FUV, USM HI peak (cyan) and cavity (yellow) contours on USM HI, and FUV and USM HI peak on HI MOM0. The $U-B$ image scale is larger than the other image scales by a factor of 1.5.

An interesting result of the $U-B$ images is that there is a large range of $U-B$ colors in the HI cavities (see also Fig. \ref{fig:UBimage_hist}). A few examples illustrate this: In CVnIdwA, most of the young regions north of the center are within the FUV peak contours, but there is also emission with intermediate age (gray tone in the figure, U-B$\sim-0.55$, $\log age \sim7.3$) in the shallow HI cavity that lies below the deep HI cavity near the center. The deep cavity is mostly old ($U-B\sim-0.1$, $\log age \sim8.1$), but also contains a few pixels with young stars. DDO 43 on the other hand has young stars only in the FUV peaks; the HI cavity slightly south of the center has $U-B\sim-0.15$, corresponding to $\log age\sim8.0$. 

DDO 46 also has stars mostly $10^8$ years or older in the cavities, but the central part of the FUV peak in the north, which is slightly offset from a USM HI peak, is also fairly old. Only the cloud peak in the north has the youngest ages, like the USM HI peak and FUV peak in the southwest.  DDO 47 also has old ages inside the elongated FUV contour to the east and west of center; the younger ages are in the small regions at the USM HI peaks. 

DDO 70 is an interesting case with a large, central FUV region covering an USM HI cavity and touching 6 USM HI peaks. The $U-B$ map has an intermediate-age population everywhere inside the low-level FUV contours and young ages at the FUV peaks. The cavity is pink in the figure, corresponding to $U-B\sim-0.3$ and $\log age \sim7.5$.

Spiral cavities in USM HI images do not show up as spirals in the $U-B$ maps, but are typically red and old, like most regions outside the active star formation. 

\section{Shear Times}
\label{sect:shear}

The $\sim10^8$ year ages of the stellar populations inside many of the cavities imply these cavities are at least that old, i.e., older than we can determine with $U-B$ colors alone.  The cavities are also relatively large compared to the galaxy disks, often comparable to the smoothing length in the USM HI image, which is half the V-band scale length. These large sizes imply large expansion times if the cavities formed by centralized pressures \citep[e.g.,][]{stewart00}. In spiral galaxies, regions that are $\sim10^8$ years old or older will shear into a trailing spiral. There are a half-dozen cases of spiral-like cavities in our dIrr sample too, but most giant cavities are more circular, and this requires low shear \citep[e.g.,][]{puche92}.

The distance over which shear distorts an object of radial extent $L$ increases with time $t$ as $LtRd\Omega/dR$ for radius in the galaxy $R$ and galaxy rotation rate $\Omega(R)$. Setting this shear distance equal to $L$ gives the time of significant shear,
$t_{\rm shear}=(Rd\Omega/dR)^{-1}$. For rotation curve $V(R)$, this is
\begin{equation}\label{sheartime}
  t_{\rm shear}=(R/V)(1-\beta)^{-1},
\end{equation}
where $\beta=d\log V/d\log R$ is the local slope of the rotation curve. 

\begin{figure*}
\begin{center}
\includegraphics[width=8cm]{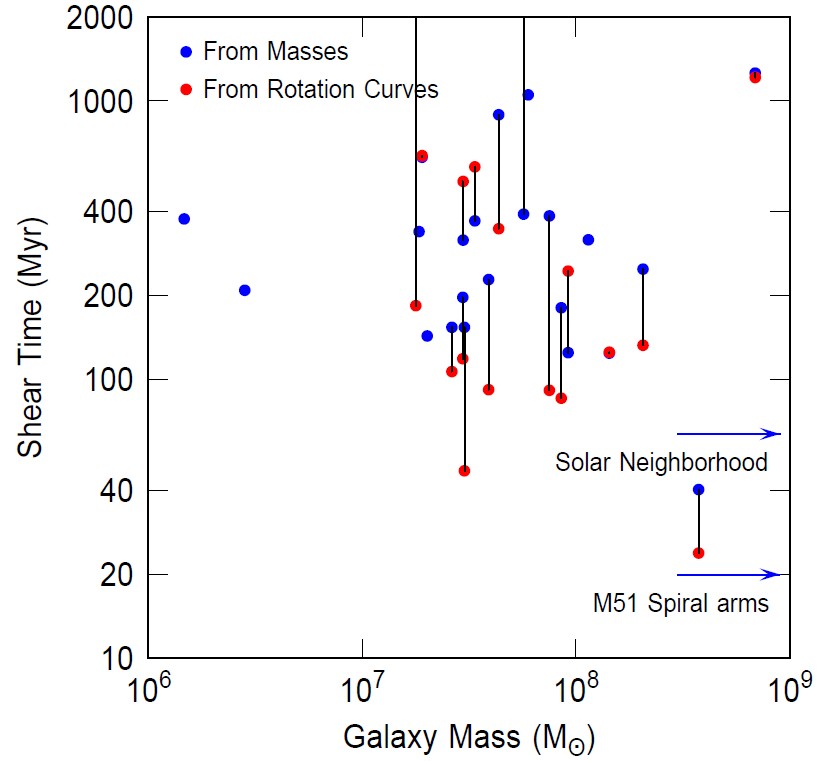} 
\caption{The time for a region to become significantly distorted as a result of rotational shear was calculated in two ways and is plotted here versus the galaxy stellar mass.  (When both methods were available for a galaxy, the points are connected by a line.) The shear time in the solar neighborhood and in a spiral arm of the grand design galaxy M51 are also plotted for comparison.  The long shear times in dIrr galaxies allow large feedback-driven cavities to remain nearly circular until long after their central pressures have subsided and the responsible stars have faded from view. 
}
\label{fig:shear}
\end{center}
\end{figure*}

Figure \ref{fig:shear} shows two evaluations of the shear time at a radius of one V-band scalelength plotted versus the galaxy stellar mass. The red points were calculated from the deprojected rotation curves of 17 LITTLE THINGS galaxies in \cite{hunter21}. The blue points were calculated from rotation curves that were derived from the deprojected distribution of mass in the form of dark matter, stars and gas. We evaluate shear times in two ways because the inclinations of dIrrs are difficult to determine, as some of these galaxies may be triaxial \citep{hunter06}.  Inclination uncertainties affect these rotation curves in different ways. For a given line-of-sight rotation speed, the deprojected speed scales as $1/\sin i$ for inclination $i$, while for a given observed surface brightness, the deprojected mass surface density scales with $\cos i$, so the deprojected rotation speed determined from the mass scales with $\sqrt( \cos i)$. 

For dark matter, we followed the derivation for dIrrs in \cite{corbelli25}, which assumed the \cite{burkert95} cored density profile, 
\begin{equation}
\rho_{\rm DM}(R) = {{\rho_0 R_0^3}\over{(R+R_0)(R^2+R_0^2)}},
\label{DM1}
\end{equation}
where $\rho_0$ and $R_0$ are the core density and radius in units of $M_\odot$ pc$^{-3}$ and pc and are related by $R_0 = 63/\rho_0$. The halo mass was obtained from the stellar mass using an abundance matching technique based on cosmological simulations \citep{dicintio14}:
$\log\rho_0=-0.175-0.21\log M_{\rm stars}$. The stellar masses are in Table \ref{properties}. 

For the stellar and gaseous contributions to the rotation curves, we integrated the accelerations to every other point in the disk as viewed from a point at radius $R$.  This is because the acceleration does not depend only on the mass interior to $R$ but on the mass distribution everywhere (unlike the halo, which is assumed to be spherical).  We used the observed, but deprojected, surface density profiles for both, converting $V$-band light and $B-V$ derived mass-to-light ratio into the stellar mass profile (see Sect. \ref{subsectiondata}) and converting HI intensity into $\Sigma_{\rm HI}$ as above (correcting for He and heavy elements).

Also in Figure \ref{fig:shear}, we show the shear time in the Solar neighborhood, which is the inverse of $2A$ for Oort constant $A=15.3$ km s$^{-1}$ kpc$^{-1}$ \citep{bovy17}. More extreme is the shear time in a spiral arm of M51, which we derive from the profiles of velocity with spiral arm phase in \cite{shetty07}. Although there are some very sharp velocity jumps in M51, many of these are shock fronts in the spiral arms and are not important for shearing cavities. Instead, we used the more gradual velocity gradients, which in the azimuthal directions are $\sim50$ km s$^{-1}$ in $\sim50^\circ$ of phase.  That $\sim50^\circ$ of phase is about one radian of angle viewed from the center, and therefore a distance of $\sim3$ kpc at mid-disk, but the important distance for the shear is the distance perpendicular to the arm, progressing into the interarm region, where the gas is moving as it shears. This distance is the $\sim3$ kpc of azimuthal distance multiplied by the tangent of the spiral arm pitch angle, which is $i=21.1^\circ$ \citep{shetty07}. Thus the local velocity gradient for cavities inside and emerging from a typical spiral arm in M51 is $\sim50$ km s$^{-1}$ divided by $\sim1$ kpc, or $\sim1/20$ Myr$^{-1}$.  The plotted shear time in Figure \ref{fig:shear} is this 20 Myr. 

Figure \ref{fig:shear} indicates that typical shear times at mid-radius in dIrrs exceed $10^8$ years and sometimes approach $10^9$ years. This easily explains the old cavities observed with red centers in $U-B$ colors, and the lack of obvious young stars. It also explains the difficulty in observing the expansions of many cavities, including the largest ones, because their sizes of a few hundred pc divided by their maximum ages of a few hundred Myr, as obtained from the shear times, are only a few km s$^{-1}$.  Cavity ages where expansion can be seen, at typical velocities of 10 km s$^{-1}$ to 15 km s$^{-1}$, have expansion ages of 20 Myr to 40 Myr \citep{pokhrel20}, which is only 10\% of the shear time. These young cavities are visible in the $U-B$ color maps where $U-B\sim-0.3$ and the color is somewhat pink (Fig \ref{colorbar}). 

\section{Properties of the HI Cavities and Peaks}
\label{sect:peaks}

The USM HI images in Figures \ref{fourtypes}, \ref{FUV-USM} and in the Appendix show cavities and peaks of HI emission after the smooth emission covering them has been removed. As indicated for Figure \ref{FUV-USM} and by equation (\ref{hicontour}), we defined the peak regions in a uniform way for all galaxies, scaled to the USM HI intensity range. In Table \ref{thresholds} and on the titles of the panels in Figures A2-A7, we give the average projected gas surface densities, $\Sigma_{\rm HI}$, inside all the $I_{\rm USM-HI,peak}$ regions for each galaxy (corrected for He and heavy elements). Here we consider other average properties of the peak HI regions to illustrate the star formation process. 

\subsection{Gas Consumption Times and the Nature of Dense Gas}

Recall that Figure \ref{SFR} shows the star formation rate surface density versus the gas surface density (without the unobserved molecules) in the HI peak regions. This turned out to be the normal relationship found elsewhere for our galaxies, as explained above. We average these averages for all the galaxies to get
\begin{equation}
{\bar\Sigma}_{\rm SFR}=0.0064\pm 0.0073 \;\;M_\odot \;{\rm pc}^{-2} \;{\rm Myr}^{-1}
\label{barsigmasfr}
\end{equation}
\begin{equation}
{\bar\Sigma}_{\rm HI}=20.2\pm 13.3 \;\;M_\odot \;{\rm pc}^{-2}
\label{eq:hiaverage}
\end{equation}
and from these, the average HI consumption time of ${\bar\Sigma}_{\rm HI}/{\bar\Sigma}_{\rm SFR}=3.16$ Gyr. The scatter in these quantities is from galaxy-to-galaxy variations, as shown by the distribution of points in Figure \ref{SFR}.  This average is comparable to the molecular consumption time in normal spiral galaxies \citep{schinnerer24}, indicating that the peak HI regions in dIrrs are somewhat like giant molecular clouds in spirals, where the metallicity is $\sim5\times$ higher and the moderately dense gas is molecular. 

We suggested previously \citep{elmegreen25} that dIrrs have average molecular fractions of $\sim0.22\pm0.1$, which would not add much to the gas mass and change this total consumption time significantly. We also suggested that the CO-rich regions in dIrrs, which tend to be very small, i.e., pc-scale inside larger HI regions \citep{hunter24}, should be viewed as the self-gravitating ``high-density'' cores of cloud complexes, much like HCN cores in spirals \citep{elmegreen25}. This conclusion is consistent with the high value of ${\bar\Sigma}_{\rm HI}$ in equation (\ref{eq:hiaverage}), which is well above the usual threshold of $\sim10\;M_\odot$ pc$^{-2}$ for the appearance of molecules in spiral galaxies \citep{bolatto11} and not unlike that of a typical CO-emitting cloud in local spiral galaxies \citep{schinnerer24}.

The pressure inside a peak HI region may be estimated from the expression $P=\pi G \Sigma_{\rm gas}^2/2$, which would be correct in the absence of stars and dark matter. Our previous paper \citep{elmegreen25} considered these contributions and showed that on average, the stellar forces contribute relatively little compared to the weight of the gas, but dark matter in the gas layer contributes a lot, effectively multiplying this simple expression by a factor of $\sim10$ (see Fig. 3 in that paper).  So, while the gas term alone would give $P=3220$ k$_{\rm B}$ cm$^{-3}$K for the average $\Sigma_{\rm HI}$ in an USM HI peak, we expect that dark matter could add another factor of a few, depending on the position in the galaxy. This total pressure would make the dynamical conditions in a typical HI peak region similar to those in the solar neighborhood. 

\subsection{Density and Scale Height from Vertical Equilibrium Solutions}

To understand more about these conditions, we need the thickness of the gas layer and the midplane density, which also gives the characteristic dynamical time.  To get both quantities, we derive the vertical equilibrium density profiles for gas and stars, including dark matter, for the 24 dIrr galaxies that were considered in our previous paper \citep{elmegreen25}.  We use the method in \cite{corbelli25}, which involves solving the equilibrium equation
\begin{equation}
c_i^2{{d\log\rho_i}\over{dz}}=-2\pi G\int_{-z}^z\left(\rho_g+\rho_s+\rho_{dm}\right)dz
\end{equation}
for velocity dispersion $c_i$ with gaseous and stellar components ``i'', density of these components, $\rho_i$, dark matter density, $\rho_{DM}$, and distance from the midplane, $z$. The dark matter density comes from equation (\ref{DM1}) and is assumed to be constant with height inside the layer. This equation was solved iteratively at each galactocentric radius for the vertical profiles of gaseous and stellar densities, constrained by the observed deprojected surface densities. We used the observed HI velocity dispersions, $\sigma_{\rm HI}$, with $c_g=1.3^{0.5}\sigma_{\rm HI}$ to account for the pressure from magnetic fields and cosmic rays, and we used a stellar velocity dispersion that scales with the stellar mass \citep[see][]{elmegreen25}.  The radial profiles of the observed quantities were determined by azimuthal averaging over annular intervals. The results of these calculations were all fairly similar for the 24 galaxies in that previous study, so we do not need to consider the additional 12 galaxies from the present paper to make the main points here.

As part of the integral, we evaluated the root mean squared density and the vertical integral of the squared density in order to estimate what type of massive star might ionize the gas completely through the average layer. The calculation gives the resulting emission measure to see if it would be observable. We also calculated the scale height as half the ratio of the observed gas deprojected surface density to the solved central gas density.

\begin{figure*}
\begin{center}
\includegraphics[width=10cm]{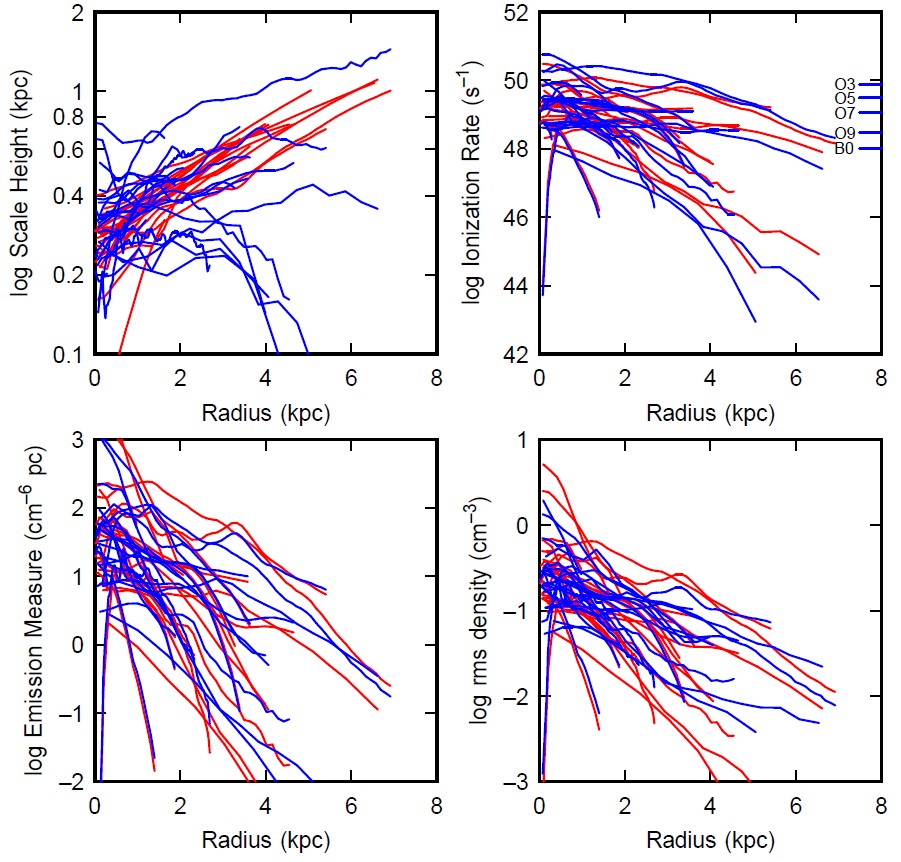} 
\caption{Results from solutions to the vertical equilibrium of gas layers with the azimuthally averaged properties of 24 dIrrs. The solutions include stellar and dark matter forces on the gas and the observed velocity dispersions. From top-left to bottom-right, the plotted quantities are the scale height, the total ionization rate needed to ionize a circular region with a diameter equal to the scale height all the way through the disk on both sides, the emission measure of such an ionized region, and the root-mean-squared density in the vertical column. Each curve is a different galaxy and all are plotted versus radius in the galaxy. The blue lines use the observed HI velocity dispersions and the red lines use an effective gas dispersion of 10 km s$^{-1}$ as if the gas were ionized. 
}
\label{fig:recomb}
\end{center}
\end{figure*}

Figure \ref{fig:recomb} shows the results plotted versus radius in each of the 24 dIrr galaxies. The blue lines in the panels use the observed HI velocity dispersion at each position for $\sigma_{\rm HI}$ and the red lines use $c_g=10$ km $^{-1}$, which might be representative of an HII region. Note first the upper left panel, which has the gas scale height for the equilibrium layer.  At mid-disk, the average scale height is $\sim400$ pc.  If we use this representative height $H$ with the average ${\bar\Sigma}_{\rm HI}$ from the USM HI peaks, we get an average midplane density in these regions of $\rho={\bar\Sigma}_{\rm HI}/(2H)=0.025\;M_\odot$ pc$^{-3}$, which corresponds to 0.76 cm$^{-3}$ of HI. The free fall time at this density is $t_{\rm ff}=(32G\rho/3\pi)^{-1/2}=50.8$ Myr.  This time combines with the average star formation rate and gas surface density to give the efficiency per free fall time, 
\begin{equation}
\epsilon_{\rm ff} = t_{\rm ff}{\bar\Sigma}_{\rm SFR}/{\bar\Sigma}_{\rm HI} = 0.016
\end{equation}
which is a typical value for spiral galaxies \citep{mckee07,krumholz19}.

The top-right panel in Figure \ref{fig:recomb} shows the total rate for Case B hydrogen recombination in the integrated equilibrium layer, considering both halves of the layer and a circular cylinder perpendicular to the plane with a diameter equal to the gas scale height.  Case B means that all recombinations from the free state to the ground state ionize another hydrogen in the layer. This recombination rate is what would balance the massive-star ionization of a significant volume of the galaxy all the way through the disk on both sides.  On the right-hand axis of the top-left panel are ticmarks showing the Lyman continuum photon rates of massive stars of various types, from \cite{sternberg03}.  The figure suggests that a few O-type stars could completely ionize a cavity at the average HI surface density for its radius, particularly at mid-radius or further out. Consequently, there could also be some escape of Lyman continuum radiation from the galaxy.  A recent study of Lyman continuum escape from the HII regions of nearby spiral galaxies is in \cite{chandar25}.

O-type stars are usually accompanied by clusters or associations of other stars with masses exceeding several thousand solar masses. Such clusters are present in dIrr galaxies \citep{cook19}. For example, in a study of five dIrr galaxies, \cite{hunter18} counted the number of clusters more massive than $10^3\;M_\odot$ and younger than 10 Myr, and the number of OB associations and O-type stars, using also the LEGUS catalog \citep{sabbi18}.  In this count, DDO 50 has 7 such clusters, 17 OB associations and 404 O-type stars; DDO 53 has 1 cluster, 11 associations and 101 O-type stars; DDO 63 has no such clusters but 6 associations and 105 O-type stars; NGC 3738 has 51 clusters, 3 associations and 281 O-type stars, and Haro 29 has 8 such clusters, 9 OB associations and 61 O-type stars. 

The formation of a $\sim10^3\;M_\odot$ cluster requires a dense gas cloud exceeding $\sim10^5\;M_\odot$, which could be any of a large number of our USM HI peaks. At the average surface density in equation (\ref{eq:hiaverage}), such a cloud  would only need to cover a region $\sim80$ pc in diameter. The average distance to our galaxies in Table \ref{properties} is 4.0 Mpc, so this diameter corresponds to a typical angular size of $4.1^{\prime\prime}$, which is comparable to or even smaller than the sizes of the USM HI peaks (recall the 1 arcmin scale bars on each image). We conclude that some of the HI cavities could have been highly ionized by star formation in their neighboring dense HI. 

The radial distributions of emission measure for hypothetical, completely-ionized parts of these galaxy disks at the {\it azimuthally-averaged} HI surface densities used for the equilibrium calculation (not the lower surface densities in the actual HI cavities) are plotted in the lower left panel of Figure \ref{fig:recomb}. They are typically in the range of 10 cm$^{-6}$ pc to 100 cm$^{-6}$ pc, with smaller values in the outer regions. From \cite{reynolds92}, the H$\alpha$ intensity and emission measure EM for case B recombination are related by
\begin{equation}
I=2.0\times10^{-18}T_4^{-0.92}\times EM\;\;{\rm erg}\; {\rm s}^{-1} {\rm cm}^{-2} {\rm arcsec}^{-1}
\end{equation}
where EM is $n_{\rm e}^2L$ for electron density $n_{\rm e}$ in cm$^{-3}$ and path length $L$ in parsecs. Setting temperature in units of $10^4$K $T_4 = 0.8$ \citep{reynolds92}, an emission measure between 10 and 100 cm$^{-6}$ pc corresponds to an H$\alpha$ intensity between $2.5\times10^{-17}$ and $2.5\times10^{-16}$ erg s$^{-1}$ cm$^{-2}$ arcsec$^{-1}$.

We estimate from the MOM0 images in Figure \ref{twoexamples} and Figures A2-A7 that HI cavities have lower HI surface densities than the azimuthally-averaged values by a factor of at least 5.  In Figure \ref{twoexamples}, the overall average intensity has a grayish color and the holes are blue, which according to the colorbar is about 1/5 of the average. 
All the MOM0 images in the Appendix are also plotted with color varying linearly with HI intensity, and the color scale shown in Figure A1, which also applies to these MOM0 images although with different values, has emission at 10\% of the peak with a light-blue color. Such colors or even bluer colors are not uncommon for the cavities in the MOM0 images, which means the average cavity is 20\% or less of the mean.  If the scale height is about the same in a cavity as in the average disk, then a 1/5 drop in HI intensity corresponds to a 1/25 drop in the emission measure and H$\alpha$ intensity of a similar cavity that is ionized.  For example, 4\% of an EM of $30$ cm$^{-6}$ pc would produce an H$\alpha$ intensity of $3\times10^{-18}$ erg s$^{-1}$ cm$^{-2}$ arcsec$^{-1}$. This is comparable to what \cite{egorov17} observe for H$\alpha$ inside a supergiant shell in Ho II (DDO 50), where they report 3 to $6\times10^{-18}$ erg s$^{-1}$ cm$^{-2}$ arcsec$^{-1}$.

The red lines in Figure \ref{fig:recomb} are about the same as the blue lines because the HI velocity dispersions from MOM2 are close to 10 km s$^{-1}$.  In that case, the pressure in an ionized cavity will not differ much from the pressure of a neutral cavity, only a factor of $\sim2$ from the larger density of particles when there are free electrons. 

The lower right panel in Figure \ref{fig:recomb} is the midplane rms density averaged over a scale height. A typical value is $\sim0.1$ cm$^{-3}$, suggesting a pressure of $\sim500k_{\rm B}$ at 5000 K for a warm neutral medium and possibly $\sim5\times$ lower in the cavities (considering the HI intensity drop discussed above). This is much lower than the internal cloud pressure from self-gravity of $\sim3220k_{\rm B}$ cm$^{-3}$K, and much higher than the thermal pressure in the center of a typical HI peak if it is cold neutral medium material. We calculated above that the average HI density in a peak is 0.76 cm$^{-3}$, so at $\sim100$K, the thermal pressure in the cloud would be $76k_{\rm B}$  cm$^{-3}$K.  The cloud pressure from gaseous self-gravity balances an internal cloud turbulence pressure at this average cloud density if the velocity dispersion is 5.1 km s$^{-1}$. 

The average HI velocity dispersion in all of the peaks in all of the galaxies is ${\bar\sigma}_{\rm HI}=11.4\pm4.4$ km s$^{-1}$. This average corresponds to the averages that determined average peak region star formation rates and HI surface densities in equations (\ref{barsigmasfr}) and (\ref{eq:hiaverage}).  Combined with the average peak density of 0.76 cm$^{-3}$, this dispersion gives a pressure of $1.63\times10^4k_{\rm B}$ cm$^{-3}$K. This is $5\times$ the pressure from gaseous self-gravity that was calculated above using the expression 
$\pi G \Sigma_{\rm gas}^2/2$, so a balance between gravitational and turbulent pressure in the HI peaks requires $\sqrt 5=2.2$ times more surface density than gas in the clouds. Presumably this excess is from a combination of disk stars and dark matter inside the clouds.

To check the importance of dark matter on the internal pressure of peak HI regions, we determine the average dark matter density for all of the galaxies at one average disk scale length using equation (\ref{DM1}).  We use the average stellar mass for our galaxies of $8.1\times10^7\;M_\odot$ in the equation for $\log\rho_0$ following equation (\ref{DM1}) and use the resulting central dark matter density, $\rho_0=0.015\;M_\odot$ pc$^{-3}$ to determine the dark matter scale length, $R_0=4.31$ kpc. The average V-band disk scale length for these galaxies is $R_{\rm D}=0.56$ kpc. These values are entered into equation (\ref{DM1}) to give a representative dark matter density at one disk scale length, $0.013\;M_\odot$ pc$^{-3}$. This dark matter density converts to an equivalent atomic density of $0.38$ cm$^{-3}$. This is about half of the atomic density inside a typical USM HI peak.  Considering that the gas would be concentrated toward the midplane while the dark matter density is more uniform, the dark matter surface density inside a typical USM HI peak should be comparable to the gas surface density, and the pressure including the dark matter would then be $\sim4\times$ the pressure from gas only, or $1.3\times10^4k_{\rm B}$ cm$^{-3}$K.  This galaxy-average gravitational-binding pressure in the USM HI peaks is close to the galaxy-average turbulent pressure using the galaxy-average HI density and velocity dispersion.

\section{Conclusions}
\label{sect:conclude}

The HI in 33 dIrr and 3 BCD galaxies studied here shows considerable structure when viewed with an unsharp mask technique. Removal of the locally smooth emission often reveals large cavities surrounded by clumpy rims of denser gas where stars form. Stars also form in scattered clouds that have no obvious connections to HI cavities.  The only exceptions are the BCD types, which differ in having more centralized FUV and HI emissions.  While the cavity/rim structure has been noted before, the apparent universality of it found here suggests a distinct star formation process that differs systematically in dIrrs from star formation in larger galaxies, where stellar dynamics play a more important role. 

dIrrs are dynamically more stable than spiral galaxies because of their lower stellar surface densities, larger thicknesses, and higher dark matter fractions. As a result, star formation in dIrrs is more controlled by stellar feedback than stellar gravity, often occurring in the high-pressure rims of what appears to be a continuous sequence of feedback-driven cavities, rather than high-pressure shocks and spurs in a sequence of spiral density waves. 

Nevertheless, inside the clouds where stars actually form, the integrated conditions and efficiencies are not very different in the two type of galaxies. If we consider on the basis of surface density and internal pressure that the HI peak regions revealed in our USM images are analogous to average molecular clouds in spiral galaxies, and that the consumption time of this HI-peak gas during star formation is comparable to that of molecular gas in spirals (Sect. \ref{sect:peaks}), then there is a direct analogy between the peak atomic clouds in dIrrs and molecular clouds in spiral galaxies. The primary difference at the cloud level is that dIrrs have low metallicities and the carbon in all but the core regions is C$^+$ \citep{cigan16,cormier19,madden20,cigan21,ramambason24}, whereas in spiral galaxies with higher metallicities and extinctions, the carbon is CO in both the cloud envelopes and the cores.  CO in the star-forming regions of dIrr galaxies is more analogous to HCN than CO in spiral galaxies with respect to region size, pressure, density contrast, and degree of self-gravity \citep{elmegreen25}.  HI densities are also lower at the same surface densities in dIrrs because the disks are thicker than in spiral galaxies, by a factor of $\sim4$, but the star formation rate densities are lower too, keeping the consumption time about the same.

The puzzle with star formation in dIrrs has been that the average surface densities and mean gravitational forces are so low that normal processes of cloud formation do not seem to work. Gas dominates stars by a factor of $\sim10$ in much of a dIrr disk, and dark matter is comparable to the gas. But now it appears that the gas surface densities in dIrrs are low only on average because of the highly partitioned structure of the gas into nearly-empty cavities and small, high-surface-density clouds between the cavities. In local regions of comparable surface densities, star formation in dIrrs is not unusual compared to the solar neighborhood, except for the lack of CO emission in dIrrs.

There is another difference between dIrrs and spiral galaxies in the local rate of shear, which is $\sim10\times$ higher in spiral galaxies. This shear makes spirals out of the star-forming regions in large galaxies, forming flocculent structure if there are no strong stellar density waves \citep{elmegreen84,elmegreen96}. In dIrrs, the shear time is much longer than the time for feedback from star formation, so the cavities and rings that feedback makes remain circular until they are destroyed by the next generation of stars that form along the rims. Low shear therefore amplifies the density irregularities of feedback-driven structure in dIrrs by not shearing it away.  

{\it Acknowledgements}
The authors are grateful to the referee for useful suggestions. 
Flagstaff sits at the base of mountains sacred to tribes throughout the region. 
We honor their past, present, and future generations, who have lived here for millennia and will forever call this place home.  D.A.H. acknowledges Lowell Observatory for publication support. GALEX was operated for NASA by the California Institute of Technology under NASA contract NAS5-98034. GALEX data presented in this article were obtained from the Mikulski Archive for Space Telescopes (MAST) at the Space Telescope Science Institute; the DOI for the paper describing the GALEX GR3 data release is 10.1086/520512.

\newpage
\begin{appendix}
\section{Additional Figures}

\setcounter{figure}{0}
\renewcommand{\thefigure}{A\arabic{figure}}

This Appendix presents images like those discussed in the main text of all 36 dIrr galaxies in our survey. The first figure shows the $U-B$ color scale that applies to all of the $U-B$ panels in the six figures that follow. For each galaxy, the panels are, from left to right, the FUV emission on a linear color scale with contours at the values given by equation (\ref{FUVcontour}); the USM HI map with cyan contours from equation (\ref{hicontour}) and yellow contours from equation (\ref{hicavity}); the MOM0 total HI image with black FUV contours and cyan USM HI peaks, and the $U-B$ color in magnitudes made from values of $-1$ in blue to 0 in red, according to the color scale in Figure A1 and with all of the contour types on it. The average HI surface density including He and heavy elements, $<\Sigma_{\rm HI}>$, inside the equation (\ref{hicontour}) contours is stated in the title of the third panel. The left-hand panel for each galaxy has a yellow scale bar indicating 1 arcmin, while the $U-B$ images have a scale that is larger than the other images by a factor of 1.5 for clarity.
 
Table \ref{thresholds} gives physical values for the black, cyan, and yellow contours in all of the figures. The last column has the average projected surface density of HI, corrected for He and heavy elements, in the regions above the USM HI thresholds given by equation (\ref{hicontour}).

Figure \ref{appendix7} shows the distributions of $U-B$ color and age for all of the galaxies.

\begin{center}
\begin{deluxetable}{lcccccc}
\tablenum{3} \tablecolumns{7} \tablewidth{280pt} 
\tablecaption{$\Sigma_{\rm SFR}$ and $\Sigma_{\rm HI}$ Thresholds \label{thresholds} } 
\tablehead{
\colhead{} & 
\colhead{$\log\Sigma_{\rm SFR,low}$\tablenotemark{a}}&
\colhead{$\log\Sigma_{\rm SFR,med}$}&
\colhead{$\log\Sigma_{\rm SFR,high}$}&
\colhead{$\Sigma_{\rm USM-HI,cavity}$}&
\colhead{$\Sigma_{\rm USM-HI,peak}$}&
\colhead{$<\Sigma_{\rm HI}>$}
}
\startdata
CVnIdwA & -2.48 & -2.15 & -1.82 & 12.1 $\pm$ 0.004 & 23.6 $\pm$ 1.8 & 25.5 \\
DDO43 & -3.03 & -2.78 & -2.54 & 5.4 $\pm$ 0.2 & 9.5 $\pm$ 0.6 & 10.2 \\
DDO46 & -2.89 & -2.63 & -2.36 & 6.7 $\pm$ 1.3 & 20.2 $\pm$ 0.9 & 22.1 \\
DDO47 & -2.71 & -2.42 & -2.12 & 3.2 $\pm$ 0.8 & 10.1 $\pm$ 0.6 & 11.0 \\
DDO50 & -2.08 & -1.69 & -1.31 & 3.0 $\pm$ 2.3 & 20.4 $\pm$ 1.9 & 23.1 \\
DDO52 & -3.14 & -2.91 & -2.67 & 2.3 $\pm$ 1.6 & 9.7 $\pm$ 0.8 & 10.1 \\
DDO53 & -2.50 & -2.17 & -1.85 & 10.3 $\pm$ 0.6 & 20.1 $\pm$ 1.4 & 22.6 \\
DDO63 & -2.68 & -2.39 & -2.09 & 4.0 $\pm$ 1.4 & 14.7 $\pm$ 2.1 & 16.1 \\
DDO69 & -2.63 & -2.33 & -2.03 & 2.2 $\pm$ 2.0 & 19.8 $\pm$ 3.4 & 23.2 \\
DDO70 & -2.11 & -1.73 & -1.35 & 3.6 $\pm$ 1.8 & 12.4 $\pm$ 1.3 & 14.1 \\
DDO75 & -2.25 & -1.89 & -1.53 & 8.1 $\pm$ 3.9 & 34.2 $\pm$ 6.0 & 39.5 \\
DDO87 & -2.85 & -2.58 & -2.31 & 1.9 $\pm$ 0.7 & 8.6 $\pm$ 0.5 & 9.61 \\
DDO101 & -3.13 & -2.90 & -2.66 & 0.46 $\pm$ 0.66 & 4.6 $\pm$ 0.4 & 4.97 \\
DDO126 & -2.65 & -2.35 & -2.05 & 6.7 $\pm$ 1.7 & 17.3 $\pm$ 1.7 & 19.0 \\
DDO133 & -2.51 & -2.19 & -1.87 & 4.3 $\pm$ 1.2 & 9.0 $\pm$ 1.8 & 11.1 \\
DDO154 & -2.65 & -2.35 & -2.05 & 8.4 $\pm$ 2.0 & 18.7 $\pm$ 1.1 & 20.4 \\
DDO167 & -2.75 & -2.46 & -2.18 & 3.2 $\pm$ 1.4 & 14.2 $\pm$ 4.1 & 18.8 \\
DDO168 & -2.55 & -2.23 & -1.91 & 10.4 $\pm$ 5.3 & 43.6 $\pm$ 3.9 & 51.1 \\
DDO187 & -2.94 & -2.67 & -2.41 & 3.7 $\pm$ 2.9 & 17.5 $\pm$ 1.6 & 19.9 \\
DDO210 & -2.75 & -2.46 & -2.17 & 3.1 $\pm$ 2.2 & 10.1 $\pm$ 1.5 & 11.0 \\
DDO216 & -3.28 & -2.97 & -2.65 & 0.55 $\pm$ 0.47 & 8.5 $\pm$ 0.5 & 9.89 \\
F564-V3 & -3.44 & -3.25 & -3.06 & 1.5 $\pm$ 0.5 & 5.1 $\pm$ 0.5 & 4.92 \\
IC1613 & -1.88 & -1.47 & -1.06 & 1.2 $\pm$ 1.3 & 16.2 $\pm$ 2.2 & 21.3 \\
LGS3 & -3.14 & -2.91 & -2.67 & 0.04 $\pm$ 0.08 & 1.3 $\pm$ 0.2 & 1.52 \\
M81dwA & -3.08 & -2.84 & -2.59 & 0.62 $\pm$ 0.35 & 4.5 $\pm$ 0.5 & 4.48 \\
NGC1569 & -2.24 & -1.88 & -1.51 & 14.0 $\pm$ 7.6 & 55.9 $\pm$ 5.5 & 65.8 \\
NGC2366 & -1.73 & -1.30 & -0.86 & 12.2 $\pm$ 5.2 & 33.7 $\pm$ 3.7 & 39.9 \\
NGC3738 & -1.96 & -1.56 & -1.16 & 7.7 $\pm$ 3.5 & 30.4 $\pm$ 1.0 & 32.7 \\
NGC4163 & -2.46 & -2.13 & -1.80 & 4.1 $\pm$ 2.1 & 11.2 $\pm$ 0.7 & 12.2 \\
NGC4214 & -1.52 & -1.06 & -0.59 & 4.0 $\pm$ 3.0 & 18.1 $\pm$ 3.8 & 20.9 \\
SagDIG & -3.16 & -2.83 & -2.49 & 0.95 $\pm$ 0.89 & 5.9 $\pm$ 1.0 & 6.85 \\
WLM & -2.22 & -1.86 & -1.49 & 2.2 $\pm$ 1.6 & 26.5 $\pm$ 1.1 & 31.5 \\
Haro29 & -1.97 & -1.57 & -1.17 & 6.9 $\pm$ 3.8 & 24.8 $\pm$ 1.9 & 25.3 \\
Haro36 & -1.88 & -1.46 & -1.05 & 7.4 $\pm$ 3.9 & 26.7 $\pm$ 2.4 & 31.8 \\
Mrk178 & -2.11 & -1.73 & -1.35 & 6.5 $\pm$ 0.1 & 12.7 $\pm$ 0.5 & 14.2 \\
VIIZw403 & -2.14 & -1.76 & -1.39 & 3.2 $\pm$ 2.6 & 18.7 $\pm$ 1.0 & 21.0 \\
\enddata
\tablenotetext{a}{$\Sigma_{\rm SFR}$ and $\Sigma_{\rm HI}$ are projected values in units of $M_\odot$ pc$^{-2}$ Myr$^{-1}$ and $M_\odot$ pc$^{-2}$. $\Sigma_{\rm USM-HI,cavity}$ and $\Sigma_{\rm USM-HI,peak}$ are the averages and standard deviations of the projected gas surface densities near the cavity and peak HI contours given by equations (\ref{hicavity}) and (\ref{hicontour}), respectively. $<\Sigma_{\rm HI}>$ is the average projected HI surface density above the peak regions in the USM images. All HI surface densities include He and heavy elements.
}

\end{deluxetable}
\end{center}

\begin{figure*}
\begin{center}
\includegraphics[width=2cm]{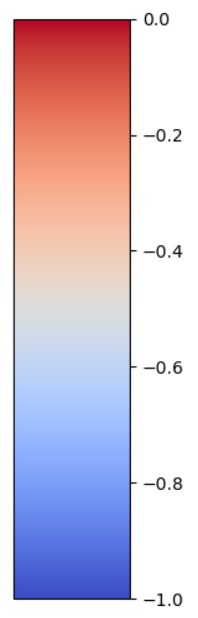} 
\caption{$U-B$ colorbar for figures A2-A7.}
\label{colorbar}
\end{center}
\end{figure*}
 
\begin{figure*}
\begin{center}
\includegraphics[width=14.7cm]{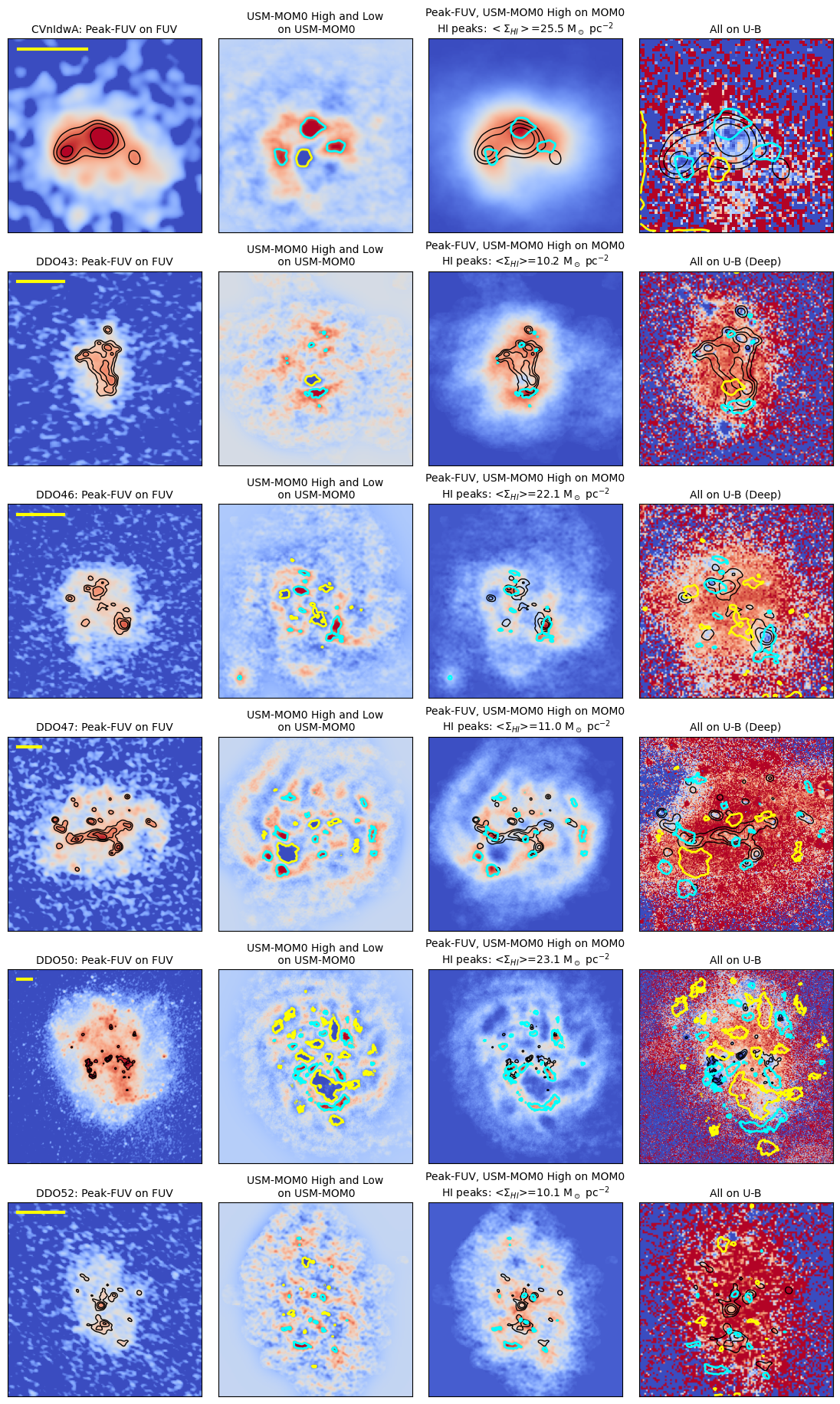}
\caption{Six galaxies displayed as in Figs. \ref{FUV-USM} and \ref{fig:UBimage}. HI and FUV-HI morphologies are listed in Table 1.
}
\label{appendix1}
\end{center}
\end{figure*}

\begin{figure*}
\begin{center}
\includegraphics[width=14.7cm]{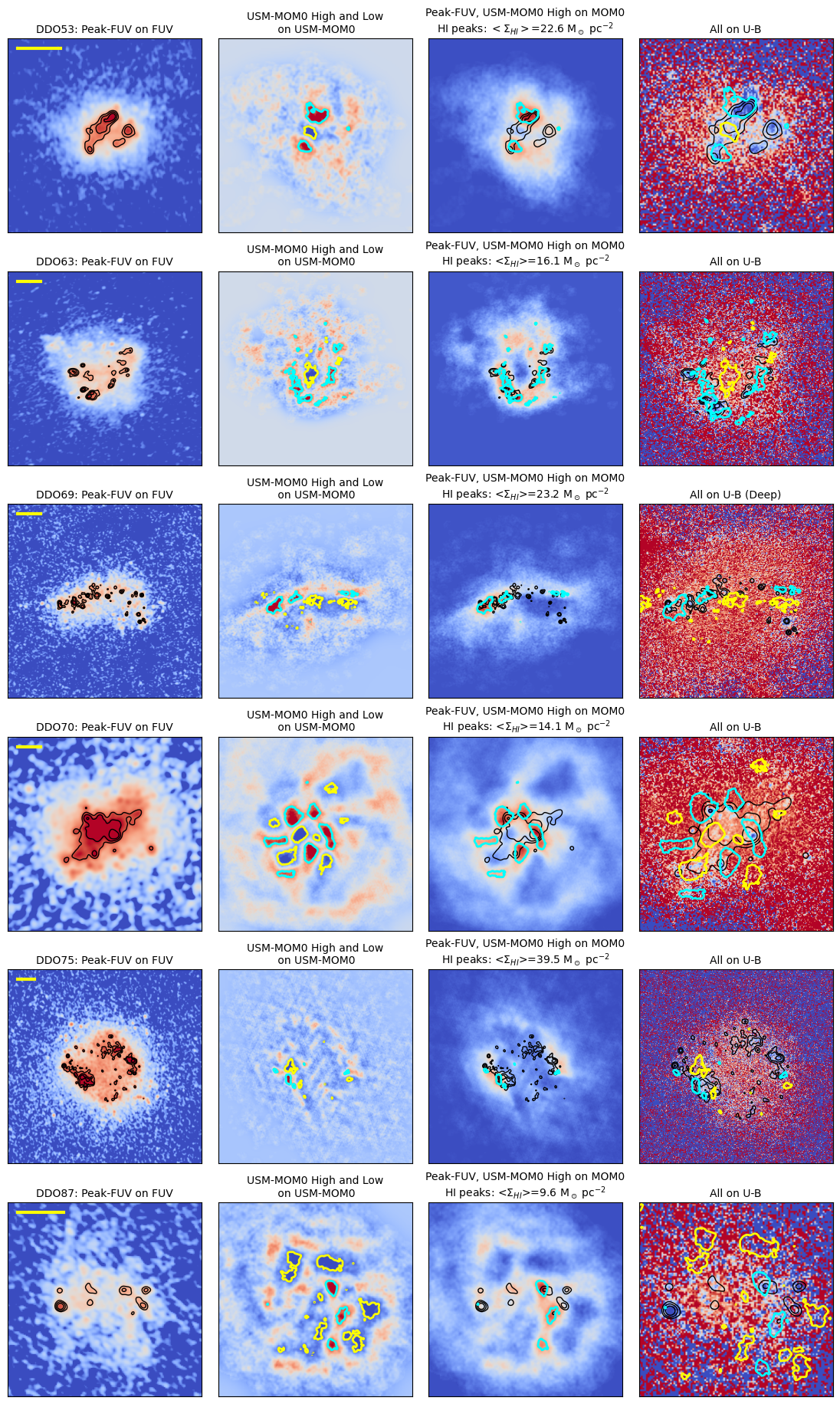}
\caption{Six galaxies displayed as in Figs. \ref{FUV-USM} and \ref{fig:UBimage}. HI and FUV-HI morphologies are listed in Table 1.
}
\label{appendix2}
\end{center}
\end{figure*}

\begin{figure*}
\begin{center}
\includegraphics[width=14.7cm]{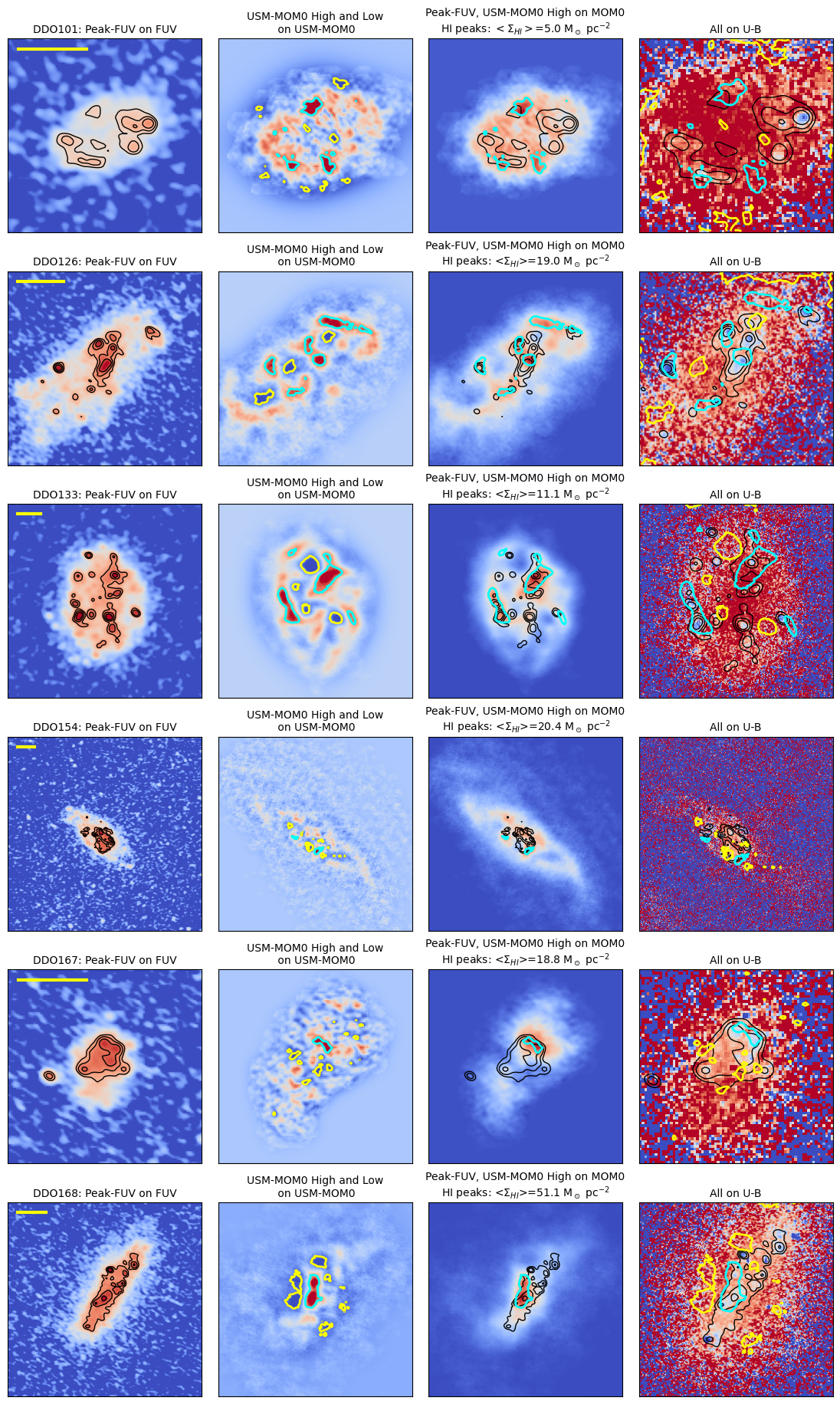}
\caption{Six galaxies displayed as in Figs. \ref{FUV-USM} and \ref{fig:UBimage}. HI and FUV-HI morphologies are listed in Table 1.
}
\label{appendix3}
\end{center}
\end{figure*}

\begin{figure*}
\begin{center}
\includegraphics[width=14.7cm]{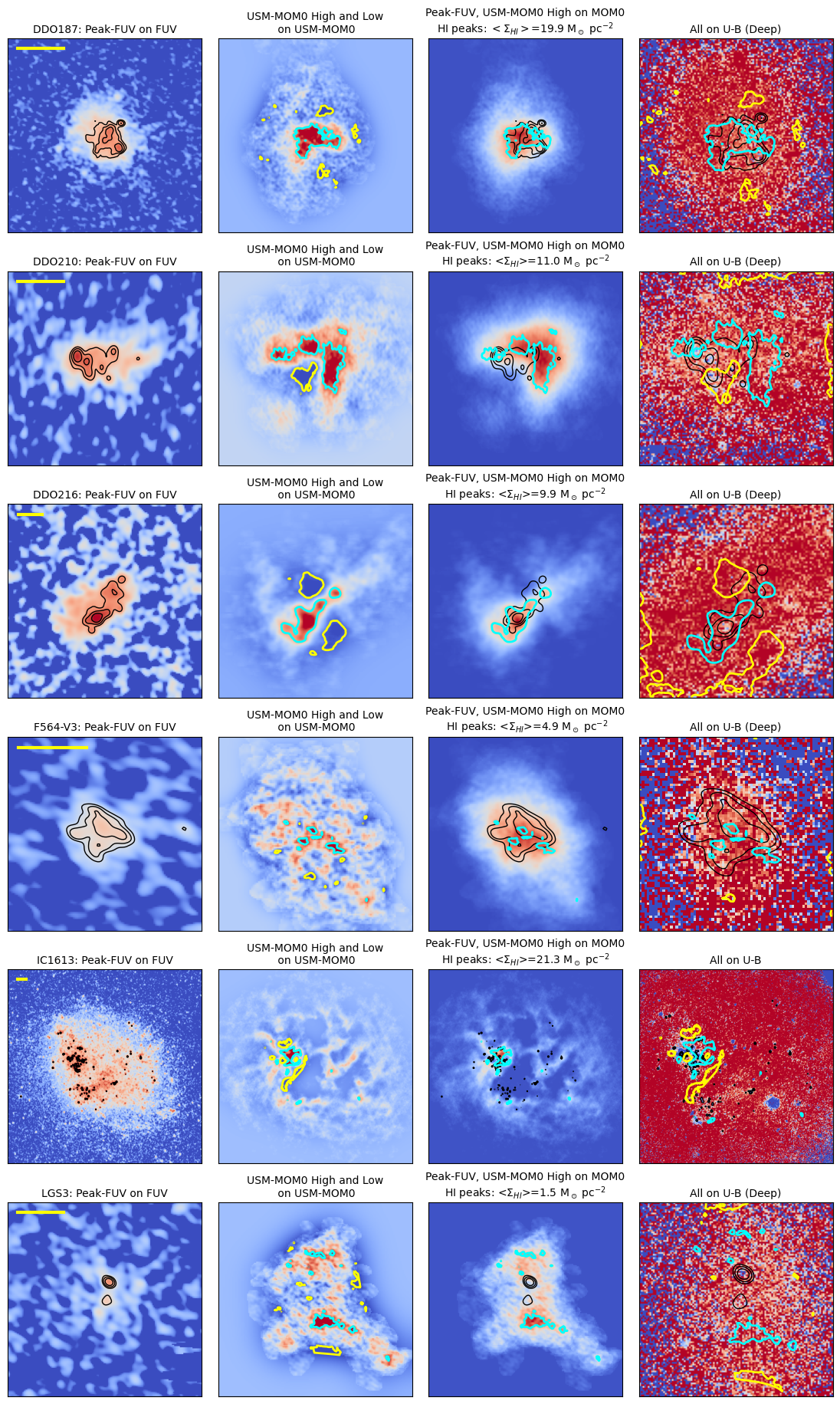}
\caption{Six galaxies displayed as in Figs. \ref{FUV-USM} and \ref{fig:UBimage}. HI and FUV-HI morphologies are listed in Table 1.
}
\label{appendix4}
\end{center}
\end{figure*}

\begin{figure*}
\begin{center}
\includegraphics[width=14.7cm]{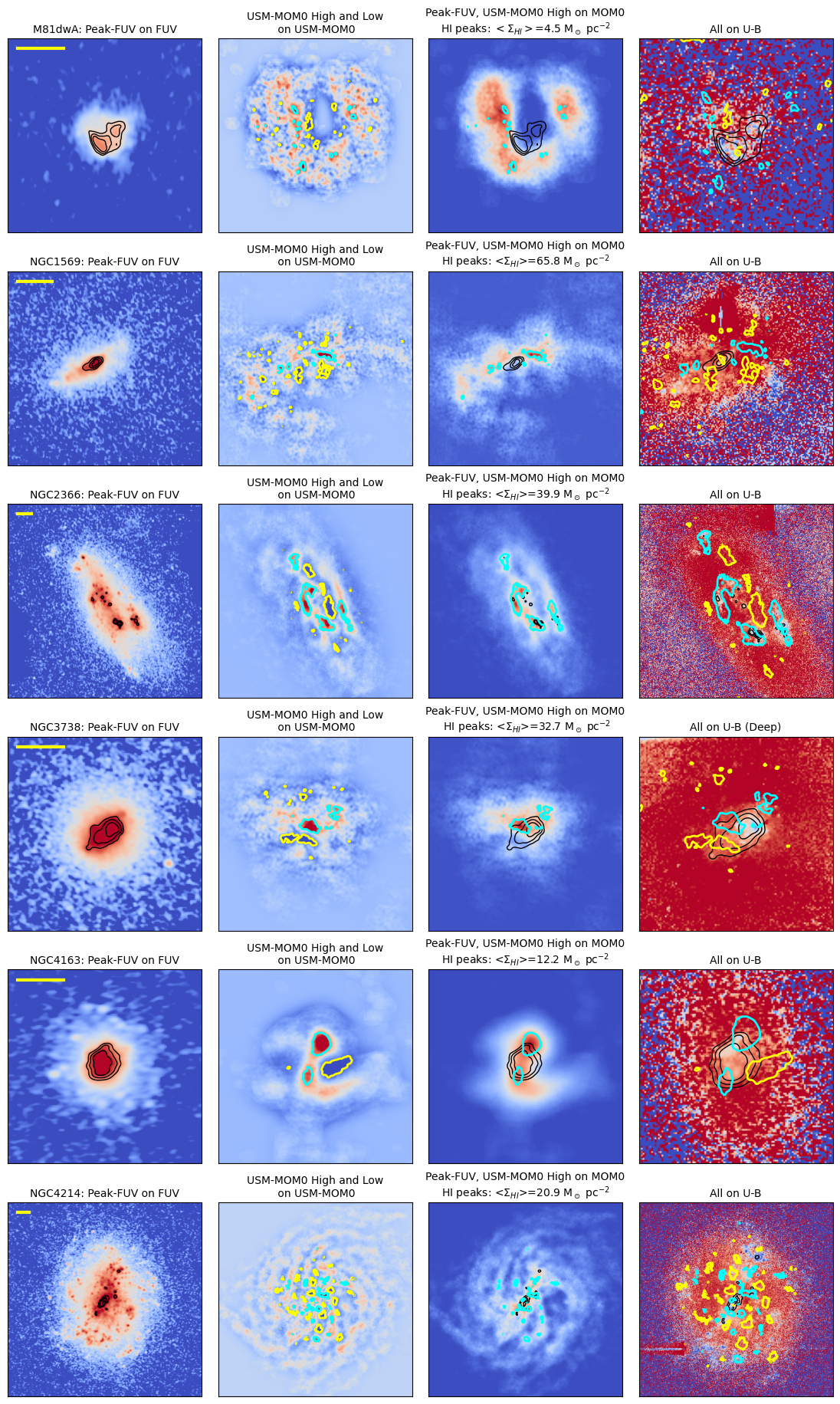}
\caption{Six galaxies displayed as in Figs. \ref{FUV-USM} and \ref{fig:UBimage}. HI and FUV-HI morphologies are listed in Table 1.
}
\label{appendix5}
\end{center}
\end{figure*}

\begin{figure*}
\begin{center}
\includegraphics[width=14.7cm]{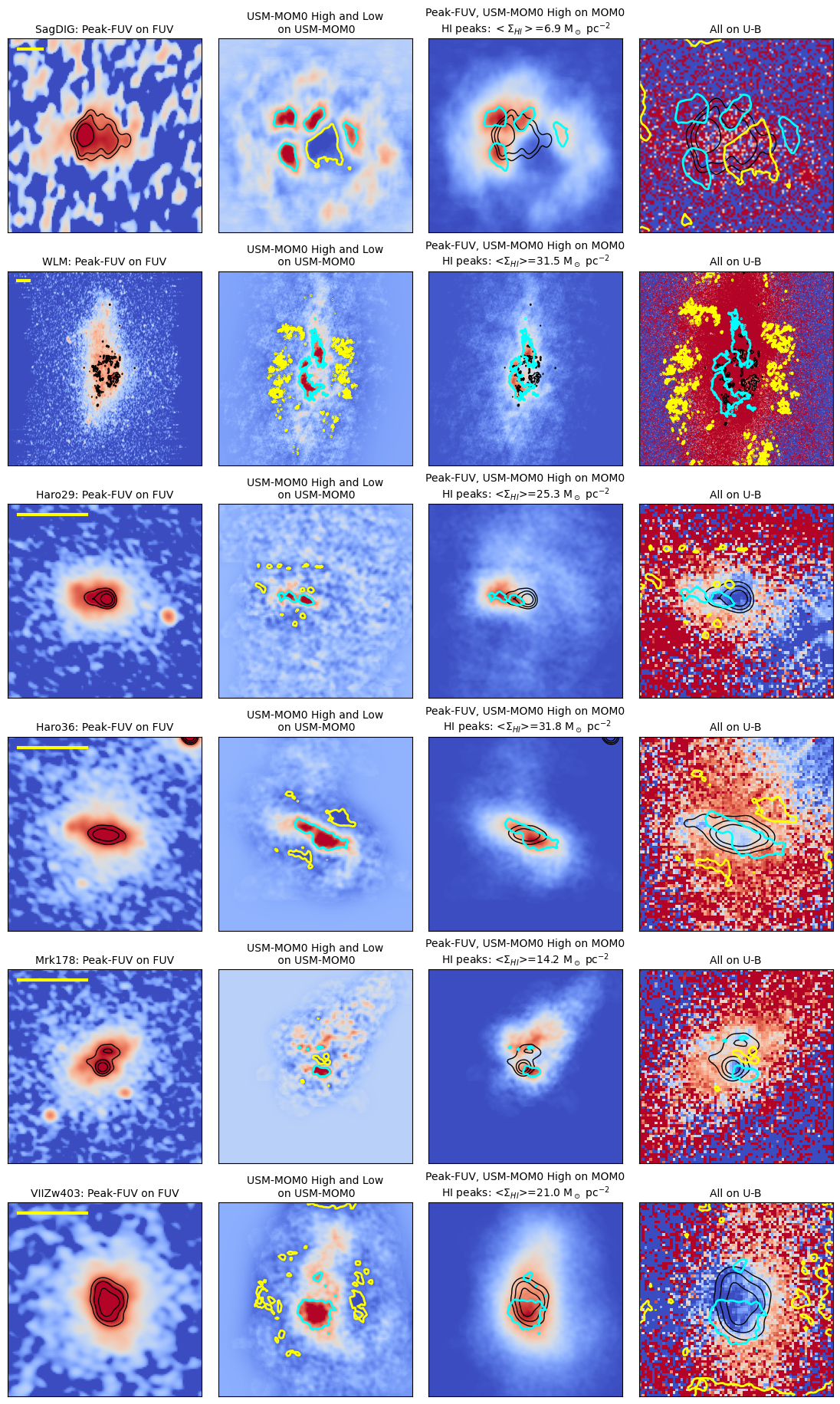}
\caption{Six galaxies displayed as in Figs. \ref{FUV-USM} and \ref{fig:UBimage}. HI and FUV-HI morphologies are listed in Table 1.
}
\label{appendix6}
\end{center}
\end{figure*}

\begin{figure*}
\begin{center}
\includegraphics[width=14.7cm]{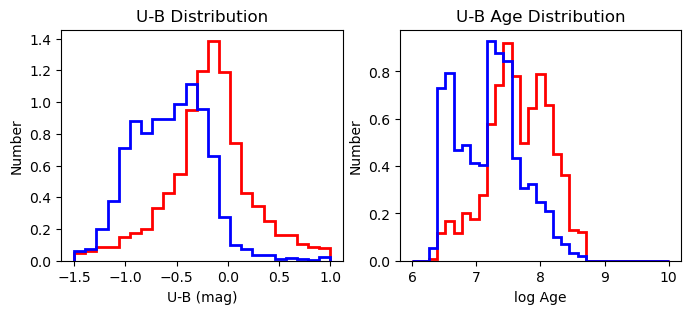}  
\caption{Histograms of color and age for all galaxies combined, with blue histograms representing the regions of peak FUV emission, as given by equation (\ref{FUVcontour}) with $A=0.9$, and the red histograms representing the cavity regions defined by equation (\ref{hicavity}). This figure is analogous to Figure \ref{fig:UBimage_hist}, which showed the color and age distributions separately for the two galaxies in Figure \ref{fig:UBimage}. Some galaxies have relatively prominent FUV emission regions which give the peak at $\log {\rm Age}\sim6.7$, and the broad plateau at $U-B\sim-1$ to $-0.5$, unlike the galaxies in Figure \ref{fig:UBimage}. 
}
\label{appendix7}
\end{center}
\end{figure*}

\end{appendix}
\newpage

\bibliography{refs}{}
\bibliographystyle{aasjournalv7}

\end{document}